\documentclass[11pt, a4paper]{article}
\pdfoutput=1
\usepackage{jheppub}
\input{epsf.tex}
\usepackage{graphics} 
\usepackage{epsfig}
\usepackage{amsfonts}
\usepackage{amsmath}
\usepackage{transparent, colortbl}
\usepackage{booktabs}
\usepackage{multirow}
\usepackage{anysize}
\usepackage{array}
\usepackage{latexsym, amscd, amsthm}
\usepackage{pdfsync}
\usepackage{epsf}
\usepackage[all]{xy}
\usepackage[usenames,dvipsnames]{xcolor}
\usepackage{enumerate}
\usepackage{float}
\restylefloat{table}
\usepackage{upgreek}
\usepackage{subcaption}
\usepackage{array}
\usepackage{multirow}
\DeclareSymbolFont{extraup}{U}{zavm}{m}{n}
\DeclareMathSymbol{\varheart}{\mathalpha}{extraup}{86}
\DeclareMathSymbol{\vardiamond}{\mathalpha}{extraup}{87}

\usepackage{fancyhdr}

\makeatletter
\def\CT@@do@color{%
  \global\let\CT@do@color\relax
        \@tempdima\wd\z@
        \advance\@tempdima\@tempdimb
        \advance\@tempdima\@tempdimc
        \kern-\@tempdimb
\transparent{0.6}%
        \leaders\vrule
                \hskip\@tempdima\@plus  1fill
        \kern-\@tempdimc
        \hskip-\wd\z@ \@plus -1fill }
\makeatother

\makeatletter
\newcommand{\thickhline}{%
    \noalign {\ifnum 0=`}\fi \hrule height 1pt
    \futurelet \reserved@a \@xhline
}
\newcolumntype{"}{@{\hskip\tabcolsep\vrule width 1pt\hskip\tabcolsep}}
\makeatother

\newtheorem{Theorem}{Theorem}[section]
\newtheorem{Lemma}[Theorem]{Lemma}
\newtheorem{Proposition}[Theorem]{Proposition}
\theoremstyle{definition}
\newtheorem{Definition}[Theorem]{Definition}
\theoremstyle{remark}

\newcommand{\bp}{\begin{Proposition}}
\newcommand{\ep}{\end{Proposition}}
\newcommand{\bl}{\begin{Lemma}}
\newcommand{\el}{\end{Lemma}}
\newcommand{\bt}{\begin{Theorem}}
\newcommand{\et}{\end{Theorem}}
\newcommand{\bd}{\begin{Definition}}
\newcommand{\ed}{\end{Definition}}
\newcommand{\End}{\mathrm{End}}

\newcommand{\eqdef}{\stackrel{{\rm def.}}{=}}

\newcommand{\cinf}{{{\cal \cC}^\infty(M,\R)}}

\DeclareFontFamily{U}{rsf}{}
\DeclareFontShape{U}{rsf}{m}{n}{<5> <6> rsfs5 <7> <8> <9> rsfs7 <10-> rsfs10}{}
\DeclareMathAlphabet\Scr{U}{rsf}{m}{n}

\newcommand{\KA}{K\"{a}hler-Atiyah~}

\def\cU{\mathcal{U}}
\def\hcU{\hat{\mathcal{U}}}
\def\cW{\mathcal{W}}
\def\hcW{\hat{\mathcal{W}}}
\def\hV{{\hat V}}
\def\hW{{\hat W}}
\def\hgamma{{\hat \gamma}}
\def\hxi{{\hat \xi}}
\def\hcB{{\hat \cB}}
\def\he{{\hat e}}
\def\hg{{\hat g}}
\def\hH{{\hat H}}
\def\hU{{\hat U}}
\def\hM{{\hat M}}
\def\hp{{\hat{p}}}
\def\hnu{{\hat \nu}}

\def\C{\mathbb{C}}
\def\R{\mathbb{R}}

\def\rk{{\rm rk}}

\def\dd{\mathrm{d}}

\def\Stab{\mathrm{Stab}}

\def\Int{\mathrm{Int}}

\def\hbcD{\hat{\cD}}
\def\hcD{\hat{\cD}}

\def\bJ{\mathbf{J}}
\def\i{\mathrm{i}}
\def\momega{{\boldsymbol{\omega}}}

\def\hOmega{\hat{\Omega}}
\def\hI{\hat{I}}
\def\hJ{\hat{J}}

\def\hphi{\hat{\varphi}}
\def\hrho{\hat{\rho}}
\def\hvarphi{\hat{\varphi}}
\def\hrho{\hat{\rho}}

\def\halpha{\hat{\alpha}}
\def\hbeta{\hat{\beta}}
\def\balpha{{\boldsymbol{\alpha}}}
\def\bn{{\mathbf{n}}}
\def\bv{{\mathbf{v}}}

\def\hJ{\hat{J}}

\def\codim{\mathrm{codim}}

\def\bJ{\mathbf{J}}
\def\bOmega{\mathbf{\Omega}}
\def\bI{\mathbf{I}}
\def\bphi{\boldsymbol{\varphi}}
\def\brho{\boldsymbol{\rho}}
\def\q{\mathfrak{q}}

\newcommand{\be}{\begin{equation*}}
\newcommand{\ee}{\end{equation*}}
\newcommand{\ben}{\begin{equation}}
\newcommand{\een}{\end{equation}}
\newcommand{\beqa}{\begin{eqnarray*}}
\newcommand{\eeqa}{\end{eqnarray*}}
\newcommand{\beqan}{\begin{eqnarray}}
\newcommand{\eeqan}{\end{eqnarray}}

\newcommand{\nn}{\nonumber}

\newcommand{\id}{\mathrm{id}}

\def\cC{{\mathcal C}}
\def\cB{\Scr B}

\def\cZ{{\cal Z}}

\def\Spin{\mathrm{Spin}}

\def\SO{\mathrm{SO}}

\def\cD{\mathcal{D}}

\def\hS{\hat{S}}

\def\cN{\mathcal{N}}

\def\cT{\mathcal{T}}

\def\cC{\mathcal{C}}

\def\SU{\mathrm{SU}}

\def\G_2{\mathrm{G_2}}

\def\cZ{\mathcal{Z}}
\def\cV{\mathcal{V}}

\def\fI{\mathfrak{I}}
\def\fD{\mathfrak{D}}

\def\fP{\mathfrak{P}}
\def\fA{\mathfrak{A}}

\title{Internal circle uplifts, transversality and stratified G-structures}

\author{Elena Mirela Babalic$^{1,2}$, Calin Iuliu Lazaroiu$^3$ }

\affiliation{
   $^1$ Department of Theoretical Physics, National
  Institute of Physics and Nuclear Engineering, Str. Reactorului
  no.30, P.O.BOX MG-6, Postcode 077125, Bucharest-Magurele, Romania  \\
 $^2$ Department of Physics, University
  of Craiova, 13 Al. I. Cuza Str., Craiova  200585, Romania\\
 $^3$ Center for Geometry and Physics, Institute for Basic Science, Pohang 790-784, Republic of Korea
}

\emailAdd{mbabalic@theory.nipne.ro, calin@ibs.re.kr} 

\abstract{We study stratified G-structures in $\cN=2$
  compactifications of M-theory on eight-manifolds $M$ using the
  uplift to the auxiliary nine-manifold $\hM=M\times S^1$. We show
  that the cosmooth generalized distribution $\hcD$ on $\hM$ which
  arises in this formalism may have pointwise transverse or
  non-transverse intersection with the pull-back of the tangent bundle
  of $M$, a fact which is responsible for the subtle relation between
  the spinor stabilizers arising on $M$ and $\hM$ and for the
  complicated stratified G-structure on $M$ which we uncovered in
  previous work. We give a direct explanation of the latter in terms
  of the former and relate explicitly the defining forms of the
  $\SU(2)$ structure which exists on the generic locus $\cU$ of $M$ to
  the defining forms of the $\SU(3)$ structure which exists on an open
  subset $\hcU$ of $\hM$, thus providing a dictionary between the
  eight- and nine-dimensional formalisms.}

\begin{document}

\maketitle 

\pagebreak

\vskip .6in

\section*{Introduction}

General $\cN=2$ flux compactifications of eleven-dimensional
supergravity \cite{sugra11} on eight-manifolds $M$ have two
independent internal supersymmetry generators $\xi_1,\xi_2$ which are
global sections of the rank sixteen bundle $S$ of Majorana spinors on
$M$. The class of such compactifications is little explored, with the
notable exception of compactifications down to Minkowski 3-space
\cite{BeckerCY}, which arise when imposing the Weyl condition on
$\xi_1$ and $\xi_2$ and which, as a consequence of no-go theorems, can
only support a flux at the quantum, rather than classical,
level. Relaxing this condition leads to backgrounds which can support
classical fluxes and which have a surprisingly rich geometry. Some
aspects of such backgrounds were discussed in \cite{Palti} using a
formalism which uses the auxiliary nine-manifold $\hM\eqdef M\times
S^1$ and the canonical lifts $\hxi_1,\hxi_2$ to $\hM$ of the internal
supersymmetry generators (see also \cite{ga2}). In that approach, one
finds that $\hM$ is endowed with a stratified G-structure whose strata
are defined by the isomorphism type of the stabilizer group inside
($\Spin(9)$) of the pair of lifted spinors at various points of
$\hM$. The strata of $\hM$ correspond \cite{Palti} to stabilizers
isomorphic with $\SU(3)$, $\G_2$ or $\SU(4)$.  On the other hand, it
was shown in \cite{msing} that the stabilizer stratification induced
by $\xi_1$ and $\xi_2$ on $M$ has $\SU(2)$, $\SU(3)$, $\G_2$ and
$\SU(4)$ strata, whose description is considerably more complex. This
stratification of $M$ coincides with a certain coarsening of the
preimage of the connected refinement of the canonical Whitney
stratification \cite{Whitney, Gibson} of a four-dimensional compact
semi-algebraic \cite{BCR, BPR} body $\fP\subset \R^4$ through a
certain map $B:M\rightarrow \R^4$ whose image is contained in
$\fP$. As shown in \cite{msing}, this complicated stratification
generalizes what happens in the much simpler case of $\cN=1$ M-theory
flux compactifications on eight-manifolds
\cite{MartelliSparks,Tsimpis,g2,g2s} (which extend the classically
fluxless case of \cite{Becker1,Becker2,Constantin}), where the
relevant semi-algebraic body is the interval $[-1,1]$, endowed with
its Whitney stratification.

The complexity of the picture found in \cite{msing} may come as a
surprise given the relative simplicity of the stabilizer
stratification of $\hM$. The purpose of this note is to explain this
difference. Embedding $M$ into $\hM$ as a hypersurface $j(M)$ located
at some fixed point of $S^1$, we show that the cosmooth \cite{Drager}
generalized distribution \cite{Freeman,BulloLewis,Michor,Ratiu} $\cD$
of \cite{msing} (which is the polar distribution defined by three
1-forms $V_1,V_2,V_3\in \Omega^1(M)$) coincides with the intersection
of $TM$ with the restriction $j^\ast(\hcD)\equiv \hcD|_{j(M)}$ of the
polar distribution $\hcD$ which is defined on $\hM$ by three 1-forms
$\hV_1,\hV_2,\hV_3\in \Omega^1(\hM)$. The latter can be expressed as
bilinears in $\hxi_1$ and $\hxi_2$. The algebraic constraints
satisfied by $V_1,V_2$ and $V_3$ as a result of Fierz identities for
$\xi_1$ and $\xi_2$ are equivalent with the algebraic constraints
satisfied by $\hV_1,\hV_2$ and $\hV_3$ as a result of Fierz identities
for $\hxi_1$ and $\hxi_2$. The intersection $\cD=j^\ast(\hcD)\cap TM$
may be pointwise transverse or non-transverse, giving rise to a
disjoint union decomposition $M=\cT\sqcup \cN$, where $\cT$ is the
{\em transverse locus} and $\cN$ is the {\em non-transverse locus} of
$M$. While $\cD$ and $j^\ast(\hcD)$ coincide when restricted to $\cN$,
the ranks of their restrictions to $\cT$ differ by one. The fact that
$\cT$ may be nonempty turns out to be responsible for the difference
between the stabilizer stratifications of $M$ and $\hM$ and explains
the increased complexity of the former when compared to the latter. In
the special case when the transverse locus is empty (which turns out
to be the case considered in \cite{Palti}), the equality
$\cD=j^\ast(\hcD)$ holds globally on $M$ and the stabilizer
stratification of $M$ is obtained directly from that of $\hM$ by
intersecting every stratum of the latter with $j(M)$. In the generic
case when $\cT\neq \emptyset$, the relation between the stabilizer
stratifications of $M$ and $\hM$ can be understood using a version of
known facts \cite{FinoTomassini, ChiossiSalamon, CabreraSU3, Knotes,
  ContiSalamon, Bedulli} regarding G-structures induced on orientable
hypersurfaces of a G-structured manifold. On the open stratum
$\cU\subset M$ which carries an $\SU(2)$ structure (the ``generic
locus'' of \cite{msing}), this observation allows one to give an
explicit formula for the defining forms of the $\SU(2)$ structure in
terms of the defining forms of the $\SU(3)$ structure which exists
\cite{Palti} on an open subset $\hcU$ of $\hM$.

The note is organized as follows. Section 1 briefly recalls some
results of \cite{msing}, to which we refer the reader for further
information. Section 2 discusses the stabilizer stratification of
$\hM$ and compares its intersection with $j(M)$ with the $B$-preimage
of the connected refinement of the canonical Whitney stratification of
$\fP$. Section 3 takes up the issue of transversality of the pointwise
intersection of $j^\ast(\hcD)$ with $TM$ and shows how the transverse
or non-transverse character of this intersection explains the
increased complexity of the stabilizer stratification of $M$ as
compared to that of $\hM$. The same section shows how the stratified
G-structure of $M$ can be obtained by reducing that of $\hM$ along
this intersection. Section 4 expresses the defining form of the
$\SU(2)$ structure which exists on the generic locus of $M$ in terms
of the defining forms of the $\SU(3)$ structure which exists on an
open subset of $\hM$, while Section 5 concludes.

\paragraph{Notations and conventions.} 
We use the same notations and conventions as reference \cite{msing},
to which we refer the reader for details. An equality which holds for
any point of a subset $A$ of a manifold is written as $=_A$.

\section{Brief summary of the eight-dimensional formalism}

Let $S$ denote the rank 16 vector bundle of Majorana spinors on $M$
(which is endowed with the admissible \cite{AC1,AC2} scalar product
$\cB$) and $\nu$ denote the volume form of $(M,g)$. Let $\gamma:\wedge
T^\ast M\rightarrow \End(S)$ be the structure morphism of $S$. Given
two Majorana spinors $\xi_1,\xi_2\in \Gamma(M,S)$ which are
$\cB$-orthonormal everywhere, we define the 0- and
1-forms\footnote{The notation $=_U$ means that a relation holds on any
  open subset $U$ of $M$ which supports a local coframe
  $(e^a)_{a=1\ldots 8}$ of $M$.}:
\beqan
\label{formsi}
&&b_i = \cB(\xi_i,\gamma(\nu)\xi_i)~,~b_3\eqdef \cB(\xi_1,\gamma(\nu)\xi_2)~\\
&&V_i =_U \cB(\xi_i,\gamma_a\xi_i)e^a~ ,~V_3 \eqdef \cB(\xi_1,\gamma_a\xi_2)e^a~~,~~W\eqdef_U \cB(\xi_1,\gamma_a \gamma(\nu)\xi_2) e^a \nn
\eeqan
with $i=1,2$ and the linear combinations:
\ben
\label{formspm3}
b_\pm\eqdef \frac{1}{2}(b_1\pm b_2)~,~V_\pm\eqdef\frac{1}{2}(V_1\pm V_2)~.
\een
It is convenient to consider the smooth map:
\be
b\eqdef (b_+,b_-,b_3):M\rightarrow \R^3~~.
\ee
The Fierz identities for $\xi_1,\xi_2$ imply \cite{msing} that
\eqref{formsi} satisfy the constraints:
\ben
\label{VWsys}
\begin{split}
&||V_-||^2+b_-^2=||V_3||^2+b_3^2~~,~~||V_+||^2+b_+^2=1-(||V_3||^2+b_3^2)\\
& \langle V_+,V_-\rangle+b_+b_-=\langle V_+,V_3\rangle+b_+b_3=\langle V_-,V_3\rangle+b_-b_3=0~~\\
&||W||^2+||V_3||^2=1+b_-^2-b_+^2~\\
&\langle W, V_+\rangle =0~~,~~ \langle W, V_-\rangle =b_3 ~~,~~\langle W, V_3\rangle =-b_-~~.
\end{split}
\een
In view of the first two relations, we define:
\ben
\label{beta_def}
\beta \eqdef \sqrt{||V_-||^2+b_-^2}=\sqrt{||V_3||^2+b_3^2}=\sqrt{1-b_+^2-||V_+||^2}:M\rightarrow \R~~.
\een
Consider the cosmooth generalized distributions:
\ben
\label{cDdef}
\cD\eqdef \ker V_+\cap \ker V_-\cap \ker V_3\subset TM~~,~~\cD_0\eqdef \cD\cap \ker W\subset \cD~~.
\een
As shown in \cite{msing}, the rank stratifications of $M$ induced by
$\cD$ and $\cD_0$ have the same open stratum, the so-called {\em
  generic locus} of $M$:
\be
\cU\eqdef \{p\in M|\rk \cD(p)=5\}=\{p\in M|\rk \cD_0(p)=4\}
\ee
while the complement $\cW\eqdef M\setminus \cU$ (the {\em non-generic locus}) decomposes as: 
\ben
\label{Wdec}
\cW=\cW_2\sqcup \cW_1\sqcup \cW_0=\cZ_2\sqcup \cZ_1\sqcup \cZ_0~~,
\een
where:
\ben
\cW_k\eqdef \{p \in \cW| \rk \cD(p)=8-k\}~~,~~\cZ_k\eqdef \{p \in \cW| \rk \cD_0(p)=8-k\}
\een
and $\cZ_3=\emptyset$. The rank stratifications of $M$ induced by $\cD$ and
$\cD_0$ are the disjoint union decompositions:
\ben
\label{Mdecfull}
M=\cU\sqcup \cW_2\sqcup\cW_1\sqcup\cW_0~~,~~M=\cU\sqcup \cZ_2\sqcup\cZ_1\sqcup\cZ_0~~.
\een
It was shown in \cite{msing} that these stratifications can be
described as different coarsenings of the $B$-preimage of the
connected refinement of the canonical Whitney stratification of a
semi-algebraic body $\fP\subset \R^4$, where $B$ is the map defined
through:
\be
B=(b,\beta):M\rightarrow \R^4~~,
\ee
a map whose image is contained in $\fP$. In particular, we have
$\cU=B^{-1}(\Int \fP)$ and $\cW=B^{-1}(\partial \fP)$, while:  
\be
\cZ_0=\cW_0~~,~~\cZ_1=\cW_1^1~~,~~\cZ_2=\cW_1^0\sqcup \cW_2~~,
\ee
where $\cW_1^0$ and $\cW_1^1$ are defined in loc. cit. and satisfy
$\cW_1^0\sqcup \cW_1^1=\cW_1$. We refer the reader to \cite{msing} for
the description of $\fP$ and of its Whitney stratification, which we
will freely use below. The description of $\cW_k$ and $\cZ_k$ as
$B$-preimages of disjoint unions of various Whitney strata of the
frontier of $\fP$ can be found in loc. cit. It was also shown in
\cite{msing} that the rank stratification of $\cD_0$ coincides with
the stabilizer stratification of $M$, whose strata are defined by the
isomorphism type of the common stabilizer group $H_p\eqdef
\Stab_{\Spin(T_pM,g_p)}(\xi_1(p),\xi_2(p))$ as $p\in M$. These
isomorphism types are $\SU(2)$, $\SU(3)$, $\G_2$ or $\SU(4)$ according
to whether $p$ belongs to $\cU$, $\cZ_2$, $\cZ_1$ or $\cZ_0$. The
stabilizer stratification is the main datum describing the
``stratified G-structure'' which is induced by $\xi_1$ and $\xi_2$ on
$M$ (see \cite{msing}).

\section{Circle uplifts to an auxiliary nine-manifold}
\label{sec:9d}

\subsection{The nine-manifold $\hM$}

\begin{figure}[H]
\begin{center}
\scalebox{0.4}{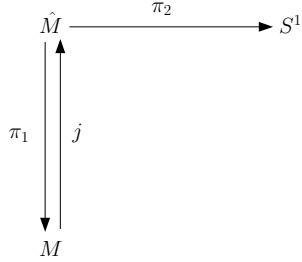}
\caption{The canonical projections $\pi_1,\pi_2$ of $\hM$ and the section $j$ of $\pi_1$.}
\label{fig:diagram}
\end{center}
\end{figure}

Following \cite{Tsimpis}, consider the 9-manifold ${\hat M}\eqdef
M\times S^1$, endowed with the direct product metric $\hg$, where
$S^1$ has unit radius. Let $s\in [0,2\pi)$ denote an angular
  coordinate on $S^1$ and $\pi_1$ and $\pi_2$ denote the canonical
  projections of $\hM$ onto $M$ and $S^1$, respectively (see Figure
  \ref{fig:diagram}). Consider the embedding $j:M\hookrightarrow {\hat
    M}$ of $M$ in ${\hat M}$ as the hypersurface given by the
  equation $s=0$:
\be
j(p)=(p,0)~~,~~\forall p\in M~~.
\ee
This gives a section of the map $\pi_1:\hM\rightarrow M$, thus
$\pi_1\circ j=\id_M$, which implies that the pull-back map
$j^\ast:\Omega(\hM)\rightarrow \Omega(M)$ satisfies $j^\ast\circ
\pi_1^\ast=\id_{\Omega(M)}$. The differential $j_\ast\eqdef \dd j:
TM\hookrightarrow T\hM|_{j(M)}$ is injective and identifies $TM$ with
the corank one sub-bundle $j_\ast(TM)$ of the restriction of $T\hM$ to
$j(M)$. To simplify notation, we identify $M$ with $j(M)$ and $TM$
with $j_\ast(TM)\subset T\hM|_{j(M)}$.
The unit circle $S^1$ is endowed with the exact one-form $\dd s$, dual
via the musical isomorphism to the Killing vector field
$\frac{\partial}{\partial s}$ which generates rotations of $S^1$. Let
$\theta\eqdef \pi_2^\ast(\dd s)=\dd (s \circ \pi_2)\in \Omega^1(\hM)$
be the normalized Killing 1-form dual to the Killing vector field
which generates $S^1$-rotations of ${\hat M}$. We orient $\hM$ by
considering the volume form:
\ben
\label{hat_nu}
\hnu \eqdef \theta\wedge \pi_1^\ast(\nu) \Longrightarrow \pi_1^\ast(\nu)=\iota_\theta{\hat \nu}~~.
\een
Notice that $\iota_{\theta}\pi_1^\ast(\nu)=0$ and that $\hnu$ is
rotationally-invariant, since so is the metric $\hg$ of $\hM$.

Let $\hS$ denote the positive signature bundle of real spinors on
$\hM$ and $\hgamma:\wedge T^\ast \hM \rightarrow \End(\hS)$ be its
structure morphism. As explained in \cite{ga2}, the vector bundle $\hS$
can be identified with the pull-back $\pi_1^\ast(S)$. The positive
signature condition means that $\hgamma(\hnu)=+\id_{\hS}$, which
amounts to:
\ben
\label{hgammatheta}
\hgamma(\theta)=\pi_1^\ast(\gamma(\nu))~~.
\een
There exists a natural $\cinf$-linear injection: 
\be
\Gamma(M,S)\ni \xi \hookrightarrow \hxi \in \Gamma(\hM,\hS)
\ee
which is constructed as explained in \cite{ga2} and whose image equals
the space of those global sections of $\hS$ which are invariant under
$S^1$-rotations of $\hM$. We say that $\hxi$ (which can be identified
with $\pi_1^\ast(\xi)$) is the {\em canonical lift} to $\hM$ of the
Majorana spinor $\xi\in \Gamma(M,S)$. The bundle $\hS$ admits a
canonical scalar product $\hcB$ which is invariant under
$S^1$-rotations of $\hM$ and hence satisfies (see \cite{ga2}):
\ben
\label{hcB}
\hcB(\hxi,\hxi')=\pi_1^\ast(\cB(\xi,\xi'))=\cB(\xi,\xi')\circ \pi_1~~,~~\forall \xi,\xi'\in \Gamma(M,S)~~.
\een

\subsection{The distribution $\hcD$}

Let $\xi_1,\xi_2$ be an everywhere-orthonormal pair of global sections
of $S$ and let $\hxi_1,\hxi_2$ be their canonical lifts to $\hM$.
Relations \eqref{hcB} show that $\hxi_1$ and $\hxi_2$ are
everywhere-orthonormal on $\hM$:
\be
\hcB(\hxi_i,\hxi_j)=\delta_{ij}~~,~~\forall i,j=1,2~~.
\ee
Consider the following one-forms defined on ${\hat M}$, where
$k=1,2,3$:
\ben
\label{hVdef}
{\hat V}_k\eqdef \pi_1^\ast(V_k)+(b_k\circ \pi_1) \theta~~,~~
{\hat V}_{\pm}=\frac{1}{2}(\hV_1\pm \hV_2)=\pi_1^\ast(V_{\pm})+(b_{\pm}\circ \pi_1) \theta~~.
\een
Relations \eqref{hgammatheta} and \eqref{hcB} imply that $\hV_k$
coincide with the natural 1-forms constructed from the canonical lifts
$\hxi_i$ of the Majorana spinors $\xi_i$:
\beqan
\label{hVspinors}
\hV_1=_\hU\hcB(\hxi_1,\hgamma_m\hxi_1)\he^m~~,~~\hV_2=_\hU\hcB(\hxi_2,\hgamma_m\hxi_2)\he^m~~,~~\hV_3=_U\hcB(\hxi_1,\hgamma_m\hxi_2)\he^m~~,
\eeqan
where $\he^m$ is any local coframe of $\hM$ defined above an open
subset $\hU\subset \hM$ and $\hgamma^m\eqdef \hgamma(\he^m)$.  The
one-forms \eqref{hVdef} are invariant under $S^1$-rotations of $\hM$,
so their Lie derivatives with respect to $\frac{\partial}{\partial s}$
vanish. Since $\pi_1^\ast(V_k)$ are orthogonal to $\theta$, we have:
\ben
\label{descent}
\langle {\hat V}_k,{\hat V}_l\rangle =(\langle V_k,
V_l\rangle +b_kb_l)\circ \pi_1~~,~~\forall k,l=1,2,3~~,
\een 
where we used the normalization property $||\theta||^2=1$. Relations
\eqref{descent} imply that the first two rows of \eqref{VWsys} are
equivalent with the following system:
\ben
\label{hVsys}
\begin{split}
&||{\hat V}_-||^2=||{\hat V}_3||^2~~,~~||{\hat V}_+||^2=1-||{\hat V}_3||^2\\
&\langle {\hat V}_+,{\hat V}_-\rangle=\langle {\hat V}_+,{\hat V}_3\rangle=\langle {\hat V}_-, {\hat V}_3\rangle=0~~
\end{split}~~,
\een
which can also be written as:
\ben
\label{hVsys2}
\begin{split}
&||{\hat V}_1||=||{\hat V}_2||=1~~,~~||{\hat V}_3||^2=\frac{1}{2}(1-\langle {\hat V}_1, {\hat V}_2\rangle)\\
&\langle {\hat V}_1,{\hat V}_3\rangle=\langle {\hat V}_2,{\hat V}_3\rangle=0
\end{split}~~.
\een
Relation \eqref{beta_def} implies: 
\ben
\label{betarel}
||\hV_-||=||\hV_3||=\sqrt{1-||\hV_+||^2}=\hbeta~~,
\een
where $\hbeta\eqdef \beta\circ \pi_1$. Relations \eqref{hVsys2}
coincide\footnote{We mention that the vector fields denoted here by
  ${\hat V}_{1,2,3}$ are denoted by $V_{1,2,3}$ in loc. cit., while
  the vector fields denoted here by ${\hat V}_\pm$ correspond to half
  of the vector fields denoted by $V_\pm$ in loc. cit., i.e. ${\hat
    V}_\pm^{\mathrm here}=\frac{1}{2}V_\pm^{\rm there}$. Compare
  \cite[eq. (2.26)]{Palti} with our relation ${\hat
    V}_\pm=\frac{1}{2}({\hat V}_1\pm {\hat V}_2)$.} with
\cite[eqs. (2.5), (2.16)]{Palti}, where they were obtained through
direct computation starting from \eqref{hVspinors} and using Fierz
identities for two spinors in nine dimensions. The common kernel of
$\hV_k$ defines a cosmooth generalized distribution on $\hM$:
\ben
\label{hcDdef}
\hcD\eqdef\ker \hV_1\cap \ker \hV_2\cap \ker \hV_3=~\ker \hV_+\cap \ker \hV_-\cap \ker \hV_3\subset T\hM~~.
\een
This distribution is invariant with respect to rotations of $\hM$.
However, notice that $\hcD$ need {\em not} be orthogonal to the
rotation generator $\theta^\sharp$ and hence it cannot be written as
the $\pi_1$-pullback of a distribution defined on $M$.

\subsection{The distribution $\hcD_0$}

One can also lift $W\in \Omega^1(M)$ to the following one-form defined
on $\hM$, which is everywhere orthogonal to $\theta$:
\ben
\label{hWdef}
\hW\eqdef \pi_1^\ast(W)=_\hU \hcB(\hxi_1,\hgamma_m\hgamma(\theta)\hxi_2)\he^m~~.
\een
The last equality follows by choosing $\he^m$ such that $\he^9=\theta$
and noticing that $\hgamma(\theta)^2=\id_{\hS}$ (since
$\theta^2=||\theta||^2=1$ in the \KA algebra of $(\hM,\hg)$) and hence
$\hcB(\hxi_1,\hgamma_9\hgamma(\theta)\hxi_2)=\hcB(\hxi_1,\hxi_2)=0$. The
system \eqref{VWsys} is equivalent with \eqref{hVsys} taken together
with the following supplementary equations:
\ben
\label{hWsys}
||\hW||^2=1+(\rho^2-\beta^2-b_+^2)\circ \pi_1~,~\langle \hW, \hV_+\rangle=0~,~\langle \hW, \hV_-\rangle=b_3\circ \pi_1~,~\langle \hW, \hV_3\rangle=-b_-\circ \pi_1~~,
\een
where:
\ben
\label{rho_def}
\rho\eqdef \sqrt{b_-^2+b_3^2}~~.
\een
The 1-forms $\hV_k$ and $\hW$ define a generalized distribution
$\hcD_0$ on $\hM$ which is rotationally-invariant:
\ben
\label{hcD0def}
\hcD_0\eqdef\hcD\cap \ker \hW\subset \hcD~~.
\een
Once again, this distribution need not be orthogonal to $\theta^\sharp$
(i.e. it need not be contained in $\ker \theta$) and hence it cannot
be written as the $\pi_1$-pullback of a distribution defined on $M$.

\subsection{The stabilizer groups for $M$ and $\hM$}

Since the natural action of $\Spin(T_\hp\hM, \hg_p)\simeq \Spin(9)$ on
$\hS_p$ induces an adjoint action on $\End(\hS_\hp)$ with respect to
which $\hgamma_m(\hp)$ transform as the components of a one-form, it
follows that the common stabilizer:
\be
\hH_\hp\eqdef \Stab_{\Spin(T_\hp\hM,\hg_\hp)}(\hxi_1(\hp),\hxi_2(\hp))~~(\hp\in \hM)~~
\ee
satisfies:
\ben
\label{hHformula}
{\hat \q}_\hp(\hH_\hp)\subset \Stab_{\SO(T_\hp\hM,\hg_\hp)}(\hV_+(\hp),\hV_-(\hp),\hV_3(\hp))~~,
\een
where ${\hat \q}_\hp:\Spin(T_\hp\hM,\hg_\hp)\rightarrow
\SO(T_\hp\hM,\hg_\hp)$ is the covering map. Notice that
$\SO(T_\hp\hM,\hg_\hp)$ does not stabilize $\theta(\hp)$. On the other
hand, the common stabilizer:
\be
H_p\eqdef \Stab_{\Spin(T_pM,g_p)}(\xi_1(p),\xi_2(p))~~(p\in M)~~
\ee
of $\xi_1(p)$ and $\xi_2(p)$ inside $\Spin(T_pM,g_p)$ satisfies
\cite{msing}:
\ben
\label{H8formula}
\q_p(H_p)\subset\Stab_{\SO(T_pM,g_p)}(V_+(p),V_-(p),V_3(p),W(p))~~(p\in M)~~,
\een
where $\q_p:\Spin(T_pM,g_p)\rightarrow \SO(T_pM,g_p)$ is the covering map.
The relation:
\be
\Stab_{\SO(T_p\hM,\hg_p)}(\theta(p))=\SO(T_pM,g_p)~~,~~\forall p\in M\equiv j(M)~~
\ee 
implies that the following holds for any point $p\in M\equiv
j(M)$:
\ben
\label{H9formula}
\Stab_{\SO(T_p\hM,\hg_p)}(\hV_+(p),\hV_-(p),\hV_3(p),\hW(p),\theta(p))\simeq \Stab_{\SO(T_pM,g_p)}(V_+(p),V_-(p),V_3(p),W(p))~~
\een
The stabilizers $\hH_\hp$ were discussed in \cite{Palti} (they can be
isomorphic with $\SU(4),\G_2$ or $\SU(3)$), while $H_p$ were computed
in \cite{msing} (they can be isomorphic with $\SU(4),\G_2,\SU(3)$ or
$\SU(2)$). As we shall see in what follows, the isomorphism type of
$\hH_\hp$ defines a stratification of $\hM$ which can be characterized
as the pull-back through a smooth and rotationally-invariant map
$\halpha\in \cC^\infty(\hM,\R)$ of the connected refinement of the
canonical Whitney stratification of a closed interval.

\subsection{The stratifications of $\hM$ and $M$ induced by $\hcD$}

\begin{figure}[H]
\begin{center}
\includegraphics[scale=0.35]{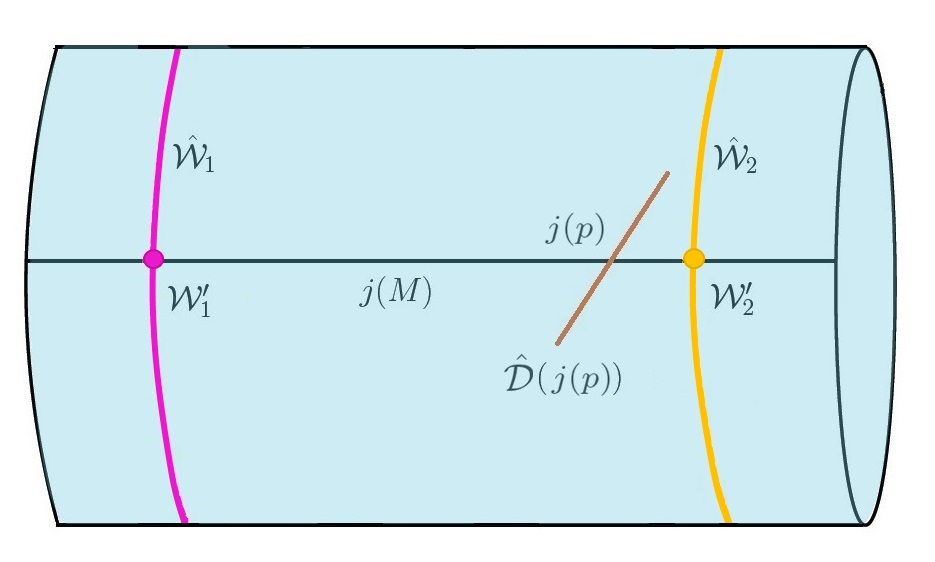}
\caption{The decomposition of $M$ induced by $\hcD$. The figure shows
  the particular case when each of the loci $\hcW_1$ and $\hcW_2$
  (depicted in magenta and yellow respectively) is connected. The open
  stratum $\hcU$ (depicted in cyan) defined by $\hcD$ is the
  complement of $\hcW=\hcW_1\sqcup \hcW_2$ inside $\hM$. The
  intersection of $\hcW_k$ with $j(M)$ determines loci $\cW'_k\subset
  M$, which in this low-dimensional rendering are depicted as
  dots. The intersection of $\hcU$ with $j(M)$ determines the locus
  $\cU'\subset M$, which is the complement of the union
  $\cW'=\cW'_1\sqcup \cW'_2$ in $M$. In brown, we depicted the space
  $\hcD(j(p))\subset T_{j(p)}\hM$ for a point $p\in M$.}
\label{fig:9d}
\end{center}
\end{figure}

The rank function of $\hcD$ gives a decomposition:
\ben
\label{hMdecomp}
\hM=\hcU\sqcup \hcW~~,
\een
where:
\ben
\hcU\eqdef \{\hp\in \hM|\rk \hcD(\hp)=6\}~~,~~\hcW\eqdef \{\hp\in \hM|\rk \hcD(\hp) > 6\}~~.
\een
The locus $\hcW$ decomposes further according to the corank of $\hcD$
inside $T{\hat M}$:
\be
\hcW=\hcW_2\sqcup \hcW_1~~.
\ee
where: 
\ben
\hcW_2\eqdef \{\hp\in \hM|\rk \hcD_\hp=7\}~~,~~\hcW_1 \eqdef \{\hp\in \hM|\rk \hcD_\hp=8\}~~.
\een
Notice that we always have $\rk \hcD(\hp)<9$, since
$||\hV_1||=||\hV_2||=1$ by \eqref{hVsys2} and hence the space spanned
by $\hV_1(\hp),\hV_2(\hp)$ and $\hV_3(\hp)$ has dimension at least
one. We thus have a disjoint union decomposition:
\ben
\hM=\hcU\sqcup \hcW_2\sqcup \hcW_1~.
\een
Also notice that $\hcU,\hcW_1,\hcW_2$ and $\hcW$ are invariant under
rotations of the circle and hence they have the forms:
\ben
\label{Mdecprime}
\begin{split}
&\hcU=\pi_1^{-1}(\cU')=\cU'\times S^1~~~~~~,~~\hcW=\pi_1^{-1}(\cW')=\cW'\times S^1~\\
&\hcW_1=\pi_1^{-1}(\cW'_1)=\cW'_1\times S^1~,~~\hcW_2=\pi_1^{-1}(\cW'_2)=\cW'_2\times S^1~\nn
\end{split}~~,
\een
where $\cU',\cW'_1,\cW'_2$ and $\cW'=\cW'_1\sqcup \cW'_2$ are subsets
of $M$ which give a decomposition (see Figure\ref{fig:9d}):
\ben
\label{Mdecprimefull}
M=\cU'\sqcup \cW'=\cU'\sqcup \cW'_2\sqcup \cW'_1~~.
\een
As we shall see below, this decompositions of $M$ induced by $\hcD$ is
generally quite different from the first decomposition in
\eqref{Mdecfull} (which is induced by $\cD$).
\noindent Using \eqref{hVsys2}, the Gram determinant formula gives:
\ben
\label{GramhV}
||\hV_1\wedge \hV_2\wedge \hV_3||^2=\det \left[\begin{array}{ccc} 1 ~& ~\halpha & 0\\ \halpha~ & ~1 &0 \\ 0 & 0 & \frac{1-\halpha}{2}\end{array}\right]=\frac{1}{2}(1+{\halpha})(1-{\halpha})^2~~,
\een
where we introduced the function (this is denoted by $\alpha$ in
\cite{Palti}):
\ben
\label{halpha_def}
{\halpha}\eqdef \langle \hV_1,\hV_2\rangle \in \cC^\infty({\hat M}, [-1,1])~~.
\een
Notice that $\halpha$ is invariant under rotations of the circle and
hence:
\be
{\halpha}=\alpha\circ \pi_1~~\mathrm{for~some~function}~~\alpha\in \cC^\infty(M,\R)~~.
\ee
Relation \eqref{GramhV} implies that the decomposition
\eqref{hMdecomp} of $\hM$ coincides with the $\halpha$-preimage of the
canonical Whitney stratification of the closed interval $[-1,1]$:
\be
\hcU=\halpha^{-1}((-1,1))=\{\hp\in \hM|~\halpha(\hp)\in (-1,1)\}~~,~~\hcW=\halpha^{-1}(\{-1,1\})=\{\hp\in \hM|~|\halpha(\hp)|=1\}~~,
\ee
while the first decomposition of $M$ given in \eqref{Mdecprimefull}
coincides with the $\alpha$-preimage of the same stratification:
\be
\cU'=\alpha^{-1}((-1,1))=\{p\in M|~\alpha(p)\in (-1,1)\}~~,~~\cW'=\alpha^{-1}(\{-1,1\})=\{p\in M|~|\alpha(p)|=1\}~~.
\ee
The following result (cf. \cite{Palti}) shows
that the rank stratification of $\hM$ induced by $\hcD$ coincides with
the $\halpha$-preimage of the connected refinement of the Whitney
stratification of the interval, while the stratification of $M$ given
by the second decomposition in \eqref{Mdecprimefull} coincides with
the $\alpha$-preimage of the same.

\paragraph{Proposition.} Let $\hp\in \hcW$.
\begin{itemize}
\item For $\halpha(\hp)=+1$, we have $\hV_3(\hp)=0$ and $\hV_1(\hp)=\hV_2(\hp)$
  with $||\hV_1(\hp)||=1$. Thus $\rk \hcD(\hp)=8$.
\item For $\halpha(\hp)=-1$, we have $\hV_2(\hp)=-\hV_1(\hp)$ with
  $||\hV_1(\hp)||=||\hV_3(\hp)||=1$ and $\hV_3(\hp)\perp \hV_1(\hp)$. Thus
  $\rk \hcD(\hp)=7$.
\end{itemize}
In particular, we have:
\beqan
\label{hcWWhitney}
\!\!\!\!&&\hcW_1=\halpha^{-1}(\{+1\})= \{\hp\in \hM|\halpha(\hp)= +1\}~,~\hcW_2=\halpha^{-1}(\{-1\})= \{\hp\in \hM|\halpha(\hp)= -1\}~~\nn\\
\!\!\!\!&&\cW'_1=\alpha^{-1}(\{+1\})= \{p\in M|\alpha(p)= +1\}~,~\cW'_2=\alpha^{-1}(\{-1\})= \{p\in M|\alpha(p)= -1\}~.~~~~~
\eeqan

\noindent{\bf Proof.} 
Follows immediately from \eqref{hVsys2}.$\blacksquare$

\

\noindent The following statement given in \cite{Palti} follows from
known facts about stabilizers of actions of Lie groups on
spheres\footnote{The stabilizer of a single non-vanishing spinor in
  the Majorana representation $\Delta_9\simeq \R^{16}$ of $\Spin(9)$
  is a subgroup isomorphic with $\Spin(7)$, belonging to a certain
  conjugacy class of subgroups of $\Spin(9)$ which is usually denoted
  by $\Spin_\Delta(7)$ (see, for example, \cite{Friedrich}). With
  respect to this subgroup, we have the decomposition
  $\Delta_9=\Lambda_7\oplus\Delta_7\oplus \R$, where $\Lambda_7\simeq
  \R^7$ and $\Delta_7\simeq \R^8$ are the vector and real spinor
  representations of $\Spin(7)$, respectively. Stabilizing
  $\hxi_1(\hp)$ first, we can take $\hxi_1(\hp)\in \R$ and
  $\hxi_2(\hp)\in \Lambda_7\oplus \Delta_7$. Thus $\hH_\hp\simeq
  \Stab_{\Spin_\Delta(7)}(\hxi_2(\hp))$ is isomorphic with
  $\SU(4)\simeq \Spin(6)$, $\G_2$ or $\SU(3)$. The first case arises
  when $\hxi_2(\hp)\in \Lambda_7$, the second when $\hxi_2(\hp)\in
  \Delta_7$ and the third when $\hxi_2(\hp)$ has non-vanishing
  projection on both $\Lambda_7$ and $\Delta_7$. In the second and
  third case, we used the fact that $\Spin(7)$ acts transitively on
  the unit sphere $S^7\subset \Delta_7$ with stabilizer $\G_2$ and the
  fact that $\G_2$ acts transitively on $S^6\subset \Lambda_7$ with
  stabilizer $\SU(3)$.}:

\paragraph{Proposition.} 
The isomorphism type of $\hH_\hp$ is given by (see Table \ref{table:9strata}):
\begin{itemize}
\item $\hH_\hp\simeq \SU(4)$ for $\hp\in \hcW_1=\halpha^{-1}(\{+1\})$
\item $\hH_\hp\simeq \G_2$ for $\hp\in \hcW_2=\halpha^{-1}(\{-1\})$
\item $\hH_\hp\simeq \SU(3)$ for $\hp\in \hcU=\halpha^{-1}((-1,1))$~~.
\end{itemize}

\begin{table}[H]
\centering {\footnotesize
\begin{tabular}{|c|c|c|c|c|c|} \hline 
$\halpha(\hp)$ & $\beta(\pi_1(\hp))$ & $B^{-1}$-stratum &
$\pi_1$-projection & $\rk \hcD(\hp)$  & $\hH_\hp$ \\ \hline 
\rowcolor{magenta}$+1$ & $0$ &$\hcW_1$ & $\cW'_1$ & $8$  & $\SU(4)$ \\ \hline 
\rowcolor{yellow}$-1$ & $1$ &$\hcW_2$ & $\cW'_2$ & $7$   & $\G_2$ \\ \hline 
\rowcolor{cyan}$\in (-1,1)$
& $\in (0,1)$ & $\hcU$ & $\cU'$ & $6$  & $\SU(3)$ \\ \hline
\end{tabular}
}
\caption{The stabilizer stratification of $\hM$. The second column of the table uses relation \eqref{hatalpha}.}
\label{table:9strata}
\end{table}

\begin{figure}[H]
\begin{center}
\includegraphics[width=0.29\linewidth]{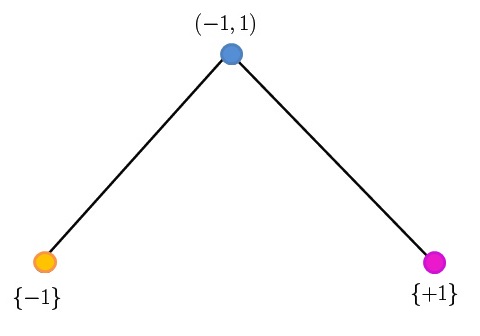}
\caption{Hasse diagram of the incidence poset (see \cite[Appendix
    C]{msing}) of the connected refinement of the Whitney
  stratification of the interval $[-1,1]$. The $\halpha$-preimages of
  the strata depicted in magenta, yellow and cyan correspond to the
  $\SU(4)$, $\G_2$ and $\SU(3)$ loci of $\hM$ respectively.}
\label{fig:WhitneyInterval}
\end{center}
\end{figure}

In particular, the stabilizer stratification of $\hM$ coincides with
the rank stratification of $\hcD$ and hence with the
$\halpha$-preimage of the canonical Whitney stratification of the
interval $[-1,1]$ (see Figure \ref{fig:WhitneyInterval}).

\subsection{Comparison with the stratification of $M$ induced by the connected refinement of the Whitney stratification of $\fP$}
\label{subsec:beta}

\noindent Let $\beta:M\rightarrow \R$ be the function defined in \eqref{beta_def}.

\paragraph{Proposition.} 
We have:
\ben
\label{hatalpha}
\alpha= \langle V_1,V_2\rangle+b_1b_2=1-2\beta^2 \in \cC^\infty(M,[-1,1])~~
\een
and hence $\beta(M)\subset [0,1]$ and:
\ben
\label{cUWprime}
\begin{split}
&\cU'=\beta^{-1}((0,1))~~,~~\cW'_1 =\beta^{-1}(\{0\})~~,~~\cW'_2 =\beta^{-1}(\{+1\})~~.
\end{split}
\een
Moreover, the following relations express $\cW'_1, \cW'_2, \cW'$ and $\cU'$ in
terms of the strata introduced in \cite[Subsection 5.3]{msing}:
\ben
\label{cWprime}
\begin{split}
&\cW'_1=B^{-1}(\fI)=\cW_0\sqcup \cW_1^0~,~~\cW'_2=B^{-1}(\fD)=\cW_1^1\sqcup \cW_2^{2+}~\\
&\cW'=\cW_0\sqcup \cW_1\sqcup \cW_2^{2+}~~,~~\cU'=B^{-1}(\Int \fP)\sqcup B^{-1}(\fA)\sqcup B^{-1}(\partial_3\fP)=\cU\sqcup \cW_2^{2-}\sqcup \cW_2^3
\end{split}~~.
\een

\noindent{\bf Proof.} Relation \eqref{hatalpha} follows from
\eqref{halpha_def} and \eqref{descent}, where the last equality in
\eqref{hatalpha} follows by subtracting the second equation of
\eqref{VWsys} from the first and using \eqref{beta_def} and
\eqref{formspm3}. Relations \eqref{cUWprime} follow immediately from
\eqref{hcWWhitney} upon using \eqref{hatalpha}. The equalities in
\eqref{cWprime} follow immediately from the last two equations in
\eqref{cUWprime} upon using the last Proposition in \cite[Subsection
  4.2]{msing} and the results of \cite[Subsection
  5.3]{msing}. $\blacksquare$

\section{Transversality}
\label{sec:t}

\subsection{Recovering $\cD$ from $\hbcD$}

To understand the relation between the rank stratifications of $\cD$
and $\hcD$, notice that \eqref{hVdef}, together with the obvious
equality $j_\ast(TM)=\ker \theta|_{j(M)}$, imply that $\cD$ can be
recovered from $\hcD$ through the relation $j_\ast
(\cD)=\left(\hcD|_{j(M)} \right)\cap j_\ast (TM)$.  Identifying $M$
with $j(M)$ (and hence $TM$ with $j_\ast(TM)$), we can write this
relation as (see Figure \ref{fig:9d}):
\ben
\label{int}
\cD=\hcD|_M\cap TM~~.
\een
Also notice that $\cU'$ and $\cW'$ coincide with the generic and
degeneration loci of the restricted distribution $\hcD|_M$:
\beqa
&&\cU'= \{p\in M|\rk \hcD(p)=6\}~~,~~\cW'= \{p\in M|\rk \hcD(p) > 6\}~~.
\eeqa

\subsection{The transverse and non-transverse loci of $M$}

Recall that two subspaces $K_1$ and $K_2$ of a vector space $K$
satisfy $\dim (K_1+K_2)=\dim K_1+\dim K_2-\dim(K_1\cap K_2)$
i.e. $\codim (K_1+K_2)=\codim K_1+\codim K_2-\codim (K_1\cap K_2)$,
where $\codim$ denotes the codimension relative to $K$. Since
$\dim(K_1\cap K_2)\leq \min(\dim K_1,\dim K_2)$, we have $\max (\codim
K_1,\codim K_2)\leq \codim (K_1\cap K_2)\leq \codim K_1+\codim K_2$.
The subspaces are called {\em transverse} when $\codim (K_1\cap K_2)=\codim
K_1+\codim K_2$, which is equivalent with $\codim(K_1+K_2)=0$
i.e. with $K_1+K_2=K$. This condition defines a symmetric binary
relation (the transversality relation) on the set of all subspaces of
$K$. For $p\in M$, let $\pitchfork_p$ denote the transversality
relation between subspaces of $T_{j(p)} \hM$, and $\not\pitchfork_p$ denote
its negation (the non-transversality relation).

\paragraph{Definition.} 
The {\em transverse locus} is the following subset of $M$:
\ben
\label{cTdef}
\cT\eqdef \{p\in M|\hcD(p)\pitchfork_p T_p M\}~~,
\een
while its complement in $M$ is called the {\em non-transverse locus}:
\ben
\label{cNdef}
\cN\eqdef \{p\in M|\hcD(p)\not \pitchfork_p T_p M\}~~,
\een
where we identify $p\in M$ with $j(p)\in \hM$ and $T_pM$ with the subspace $j_{\ast,p}(T_pM)$ of $T_{j(p)}\hM$. 

\subsection{Characterizing the transverse and non-transverse loci}
\label{subsec:TNchar}

\paragraph{Proposition.} 
Let $p\in M\equiv j(M)$. Then: 
\ben
\label{dimensions}
\dim \cD(p)\in \{\dim \hcD(p), \dim\hcD(p)-1\}~~.
\een
Moreover, the following statements are equivalent:
\begin{enumerate}[(a)]
\item $p\in \cN$
\item $\dim \cD(p)=\dim \hcD(p)$ 
\item $\cD(p)=\hcD(p)$
\item $\hcD(p) \subset T_p M$
\item $\theta(p)\in \langle \hV_1(p),\hV_2(p),\hV_3(p)\rangle$~.
\end{enumerate}
In particular, we have $\dim \cD(p)=\dim \hcD(p)-1$ iff
$p\in \cT$.

\

\noindent{\bf Proof.} 
Since $T_p{\hat M}$ has dimension nine while $T_p M$ has dimension
eight (thus $\codim T_pM=1$), relation \eqref{int} implies
$\codim\cD(p)\leq \codim \hcD(p)+1$, i.e. $\dim \cD(p)\geq \dim
\hcD(p)-1$, with equality iff $\hcD(p)$ and $T_pM$ are transverse
inside $T_p\hM$. Since $\cD(p)=\hcD(p)\cap T_pM$, we have $\dim
\cD(p)\leq \dim \hcD(p)$. This gives \eqref{dimensions} and shows
that:
\be
\hcD(p)\pitchfork T_pM~~\mathrm{iff}~~ \dim\cD(p)=\dim \hcD(p)-1~~.
\ee
The non-transverse case corresponds to $\dim \cD(p)=\dim\hcD(p)$, which
is equivalent with $\cD(p)=\hcD(p)$ since $\cD(p)$ is a subspace of
$\hcD(p)$. Since $\cD(p)=\hcD(p)\cap T_pM\subset T_pM$, the equality
$\cD(p)=\hcD(p)$ holds iff $\hcD(p)\subset T_pM$. Since $T_pM=\ker
\theta(p)$ and $\hcD(p)=\cap_{i=1}^3 \ker \hV_i(p)$, this happens iff
$\cap_{i=1}^3 \ker \hV_i(p)\subset \ker \theta(p)$, which by duality
(taking polars) happens iff $\theta(p)\in \langle
\hV_1(p),\hV_2(p),\hV_3(p)\rangle$. $\blacksquare$

\paragraph{Corollary.} 
Let $p\in M\equiv j(M)$. Then the following statements are equivalent:
\begin{enumerate}[(a)]
\item $p\in \cN$. 
\item There exist $\lambda_1,\lambda_2,\lambda_3\in \R$ such that: 
\be
\lambda_1 V_1(p)+\lambda_2 V_2(p)+\lambda_3 V_3(p)=0~~\mathrm{and}~~\lambda_1 b_1(p)+\lambda_2 b_2(p)+\lambda_3 b_3(p)=1~~.
\ee
In particular, the non-transverse locus is contained in the
degeneration locus of $\cD$ and hence the generic locus of $\cD$ is
contained in the transverse locus:
\ben
\label{NTinclusions}
\cN\subset \cW~~,~~\cU\subset \cT~~.
\een
\end{enumerate}

\noindent{\bf Proof.} 
Follows immediately from \eqref{hVdef} and from the
characterization of non-transversality given at point (e) of the
previous proposition, using the fact that $\theta(p)$ is orthogonal to
$V_k(p)$. $\blacksquare$

\subsection{Expressing $\cT$ and $\cN$ through the preimage of the connected refinement of 
the Whitney stratification of $\fP$}

\paragraph{Proposition.} 
The transverse and non-transverse loci are given by the following
unions of the strata introduced in \cite[Subsection 5.3]{msing}:
\ben
\cT=\cW_1^0\sqcup \cW_2^{2+}\sqcup \cU~~,~~\cN=\cW_0\sqcup \cW_1^1\sqcup \cW_2^{2-}\sqcup \cW_2^3~~
\een
and we have the relations:
\ben
\begin{split}
&\cU'\cap \cT=~~\cU~~~~,~~\cU'\cap\cN=\cW_2^{2-}\sqcup \cW_2^3~~\\
&\cW'_1\cap \cT=\cW_1^0~~,~~\cW'_1\cap \cN=\cW_0\\
&\cW'_2\cap \cT=\cW_2^{2+}~,~\cW'_2\cap \cN=\cW_1^1~~.
\end{split}
\een

\noindent{\bf Proof.} 
Follows immediately by comparing the ranks of $\hcD|_M$ and $\cD$ on
various loci and using relations \eqref{cWprime}, the characterization
of non-transversality given in the previous subsection and the results
summarized in Tables 5 and 6 of \cite{msing}. $\blacksquare$

\

\noindent The situation is summarized in Table \ref{table:transversality}. 
\begin{table}[H]
\centering
{\footnotesize
\begin{tabular}{|c|c|c|c|c|c|c|c|c|c|}
\hline $\fP$-locus & $\hcD$-stratum & $B^{-1}$-stratum & $\cD_0$-stratum & $\rk \hcD$ & $\rk \cD$ &  $\rk\cD_0$   & transversality & $\hH_p$ &  $H_p$\\
\hline
\rowcolor{magenta}$\partial_0\fP=\partial \fI$  &$\cW'_1$ & $\cW_0$ & $\cZ_0$ & $8$ & $8$  &  $8$   & $\cN$ & $\SU(4)$  & $\SU(4)$ \\
\hline
\rowcolor{cyan}$\partial_1^0\fP=\Int \fI$  &$\cW'_1$ & $\cW_1^0$  & $\cZ_2$ & $8$ & $7$  & $6$ & $\cT$  & $\SU(4)$ & $\SU(3)$ \\
\hline
\rowcolor{yellow}$\partial_1^1\fP=\partial \fD$  &$\cW'_2$ & $\cW_1^1$ & $\cZ_1$ & $7$ & $7$  & $7$  & $\cN$  & $\G_2$  & $\G_2$ \\
\hline
\rowcolor{cyan}$\Int \fD\subset \partial_2\fP$ &$\cW'_2$ & $\cW_2^{2+}$  & $\cZ_2$ & $7$ & $6$  & $6$   & $\cT$  & $\G_2$ & $\SU(3)$\\
\hline
\rowcolor{cyan}$\fA\sqcup \partial_3\fP$  & $\cU'$ & $\cW_2^{2-}\sqcup \cW_2^3$  & $\cZ_2$ & $6$ & $6$  & $6$  & $\cN$ & $\SU(3)$ & $\SU(3)$  \\
\hline
$\Int\fP$ & $\cU'$ & $\cU$ & $\cU$ & $6$ & $5$  & $4$  & $\cT$ & $\SU(3)$ & $\SU(2)$\\\hline
\end{tabular}
}
\caption{The ranks of $\hcD|_M, \cD$ and $\cD_0$ on various loci of $M$ and
  the character of the intersection $\hcD|_M\cap TM$. The stabilizer
  groups on $\hM$ and $M$ are shown in the last two columns.}
\label{table:transversality}
\end{table}

\paragraph{Remark.} The proposition implies: 
\be
\begin{split}
&\cU=\{p\in M |\rk \hcD(p)=6~and~\hcD(p)~\mathrm{intersects}~T_pM~\mathrm{transversely}\}=\cU'\cap \cT\\
&\cW=\{p\in M|\rk \hcD(p)>6~~or~~\hcD(p)~\mathrm{intersects}~T_pM~\mathrm{non-transversely}\}=\cW'\cup \cN~~.
\end{split}
\ee
In particular, the $\SU(2)$ stratum $\cU$ of $M$ is the intersection
of the $\SU(3)$ stratum $\hcU$ of $\hM$ with the locus $j(\cT)\subset
j(M)$, while the degeneration points of $\cD$ (the points of the locus
$\cW\subset M$) are of three kinds:
\begin{itemize}
\itemsep 0.0em
\item The points $p\in \cW'_1=\cW_0\sqcup \cW_1^0$ (where $\beta=0$
  i.e. $\alpha=+1$), which form the intersection of the $\SU(4)$
  stratum $\hcW_1$ of $\hM$ with $j(M)$. At such points, we have
  $H_p\simeq \SU(4)$ or $\SU(3)$ according to whether $p\in \cN$ or
  $p\in \cT$.
\item The points of $\cW'_2=\cW_1^1\sqcup \cW_2^{2+}$ (where $\beta=1$
  i.e. $\alpha=-1$), which form the intersection of the $\G_2$ stratum
  $\hcW_2$ of $\hM$ with $j(M)$. At such points, we have $H_p\simeq
  \G_2$ or $\SU(3)$ according to whether $p\in \cN$ or $p\in \cT$.
\item The points of $\cW\setminus \cW'=\cW_2^{2-}\sqcup \cW_2^3$,
  which form the intersection of the $\SU(3)$ stratum $\hcU$ of $\hM$
  with the locus $j(\cN)\subset j(M)$. At such points, we have
  $H_p\simeq \SU(3)$.
\end{itemize}

\subsection{The case $\cT=\emptyset$}

The previous proposition immediately implies the following:

\paragraph{Corollary.} 
The condition $\cT=\emptyset$ is equivalent with the conditions
$\cW_1^0=\cW_2^{2+}=\cU=\emptyset$. When this condition is satisfied,
we have $\cW'_1=\cZ_0=\cW_0$, $\cW'_2=\cZ_1=\cW_1^1$ and
$\cU'=\cZ_2=\cW_2^{2-}\sqcup \cW_2^3$. In this case, we have
$M=\cN=\cW_0\sqcup \cW_1^1\sqcup \cW_2^{2-}\sqcup \cW_2^3$ and
$\hH_p\simeq H_p$ for any $p\in M$, both groups being isomorphic with
$\SU(4)$, $\G_2$ or $\SU(3)$ according to whether $p\in \cW_0$, $p\in
\cW_1^1$ or $p\in \cW_2^{2-}\sqcup \cW_2^3$.

\

\noindent Notice that $\cT=\emptyset$ implies
$B^{-1}(\Int\fP)=\cU=\emptyset$ and hence requires that the image of
$B$ be contained in the frontier $\partial \fP$ of $\fP$. More precisely, 
we have: 
\be
\cT=\emptyset~~\mathrm{iff}~~B(M)\subset \partial \fI\sqcup \partial \fD\sqcup \fA\sqcup \partial_3\fP~~.
\ee
\paragraph{Remark.} 
Reference \cite{Palti} uses the assumption (see equation (3.9) of
loc. cit.)  that $\theta(p)$ is a linear combination of
$\hV_1(p),\hV_2(p)$ and $\hV_3(p)$ for {\em every} point $p\in M$. By
the characterization given at point (e) of the Proposition of
Subsection \ref{subsec:TNchar}, this assumption is equivalent with the
requirement that the transverse locus $\cT$ be empty and hence that we
are in the setting of the Corollary above. By the Corollary of
Subsection \ref{subsec:TNchar}, the condition $\cT=\emptyset$ requires,
in particular, that the 1-forms $V_1(p),V_2(p)$ and $V_3(p)$ be
linearly dependent at every point $p\in M$ (cf. \cite[Appendix
  G]{msing}). In was shown in \cite{msing} that, generically, we have
$\cU\neq \emptyset$ and hence the transverse locus is not empty in the
generic case.

\subsection{Relation between the stabilizer stratifications of $M$ and $\hM$}

It is known that an orientable hypersurface in an 8-manifold with
$\SU(4)$ structure carries a naturally induced $\SU(3)$ structure
(see, for example, \cite[Section 4]{FinoTomassini}). An orientable
hypersurface in a 7-manifold with $\G_2$ structure carries a naturally
induced $\SU(3)$ structure (see, for example, \cite{CabreraSU3,
  ChiossiSalamon, Knotes}). Finally, an orientable hypersurface of a
manifold with $\SU(3)$ structure carries a naturally induced $\SU(2)$
structure \cite{ContiSalamon}. Since these statements are purely
algebraic, they extend immediately to the case of Frobenius
distributions. Using these facts and the results above, we can
understand how the stratified G-structure of $\hM$ induces the
stratified G-structure of $M$. Namely, we have (see Table
\ref{table:transversality}):
\begin{itemize}
\item The restriction $\cD|_\cN$ coincides with $\hcD|_\cN$ and hence
  $\cD|_\cN$ carries the same structure group (namely $\SU(4)$, $\G_2$
  or $\SU(3)$) as $\hcD|_\cN$ on the components $\cW_0$, $\cW_1^1$ and
  $\cW_2^{2-}\sqcup \cW_2^3$ respectively of the non-transverse locus.
\item The restriction $\cD|_\cT$ is an orientable and corank one
  generalized sub-distribution of $\hcD|_\cT$ and hence $\cD|_\cT$
  carries the structure group $\SU(3)$, $\SU(3)$ and $\SU(2)$ on the
  components $\cW_1^0$, $\cW_2^{2+}$ and $\cU$ respectively 
 of the transverse locus $\cT$ on which
  $\hcD|_\cT$ has the structure group $\SU(4)$, $\G_2$ and $\SU(3)$
  respectively.
\end{itemize}
\noindent These observations give a different way to understand the
results of \cite{msing}, provided that one knows the codimension of
$\cD(p)$ inside $\hcD(p)$ on the various strata (which follows from
loc. cit.).

\section{Explicit relation between the $\SU(3)$ structure on $\hcU$ 
and the $\SU(2)$ structure on $\cU$}

Since $j$ identifies $\cU=\cU'\cap \cT$ with $\hcU\cap j(\cT)$, the
restriction of $\hcD$ to the locus $\cU\equiv j(\cU)\subset \hcU$ is a
regular Frobenius distribution of rank six. Since ${\hat M}$ is
oriented with volume form \eqref{hat_nu}, we can orient $\hcD|_\cU$
using the volume form:
\ben
\label{hnuperp}
\hnu_\perp \eqdef \frac{1}{||\hV_+\wedge \hV_-\wedge
 \hV_3||}\iota_{\hV_+\wedge \hV_-\wedge \hV_3}\hnu|_{\cU}~~.
\een

\subsection{The projection of $\theta$ along $\hcD$ on the generic locus}

\begin{figure}[H]
\begin{center}
\includegraphics[scale=0.5]{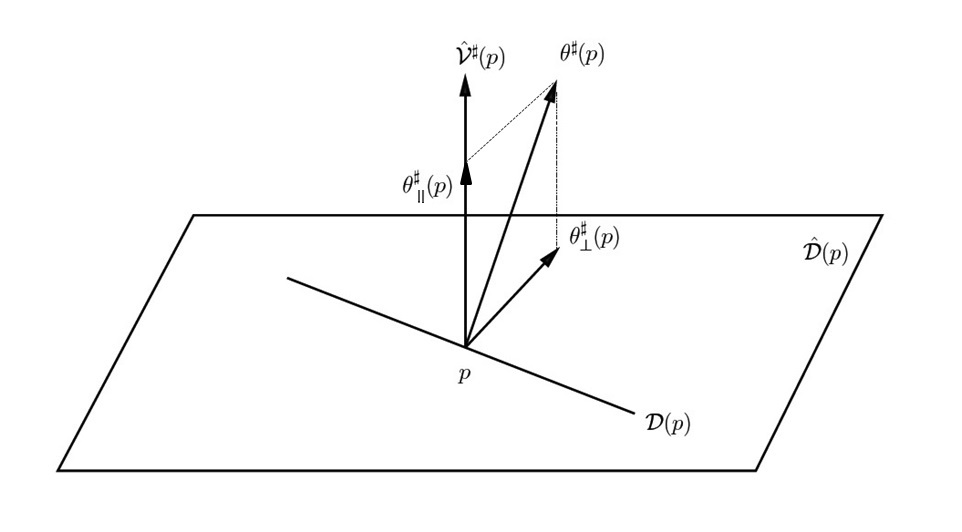}
\caption{Construction of $\theta_\perp(p)$ for $p\in \cU\equiv
  j(\cU)\subset j(M)\subset \hM$.  The vectors $\theta^\sharp(p)\in
  T_p\hM$ and $\theta_\perp^\sharp(p)\in T_p\hM$ shown in the figure
  are obtained by applying the musical isomorphism of $(\hM,\hg)$ to
  the 1-forms $\theta(p)$ and $\theta_\perp(p)$. The vertical arrow
  represents the space ${\hat \cV}^\sharp(p)$ spanned by the vectors
  $\hV_1^\sharp(p),\hV_2^\sharp(p)$ and $\hV_3^\sharp(p)$ inside
  $T_p\hM$. The vectors $\theta_\perp^\sharp(p)$ and
  $\theta_\parallel^\sharp(p)$ are the orthogonal projections of
  $\theta^\sharp(p)$ onto $\hcD(p)$ and ${\hat \cV}^\sharp(p)$
  respectively. The subspace $\cD(p)$ of $T_p\hM$ coincides with the
  intersection of $\hcD(p)$ with $T_pM=\ker\theta(p)\subset T_p\hM$
  and hence it is orthogonal to the vector $\theta^\sharp(p)\in
  T_p\hM$. It is also orthogonal to the subspace ${\hat
    \cV}^\sharp(p)\subset T_p\hM$.}
\label{fig:theta}
\end{center}
\end{figure}

The one-form $\theta|_\cU$ decomposes uniquely as:
\ben
\label{theta_dec}
\theta|_\cU=\theta_\perp+\theta_\parallel~~,
\een
where $\theta_\perp\in \Omega^1_{\cU}(\hcD)=\langle
\hV_+|_\cU,\hV_-|_\cU,\hV_3|_\cU\rangle^\perp $ and $\theta_\parallel
\in \Omega^1_{\cU}(\hcD^\perp)=\langle
\hV_+|_\cU,\hV_-|_\cU,\hV_3|_\cU\rangle$ (see Figure
\ref{fig:theta}). Since $\cU$ is a subset of $\cT$, the
characterization at point (e) of the Proposition of Subsection
\ref{subsec:TNchar} gives $\theta|_\cU \not\in \langle
\hV_+|_\cU,\hV_-|_\cU,\hV_3|_\cU\rangle$ and hence $\theta_\perp\neq
0$ and we can define the unit norm one-form:
\ben
\label{n_def}
\bn\eqdef \frac{\theta_\perp}{||\theta_\perp||}\in\Omega^1_\cU(\hcD)~~.
\een
We orient the rank five Frobenius distribution $\cD|_\cU$
such that its volume form is given by:
\ben
\label{nuperp}
\nu_\perp=-\frac{1}{||V_+\wedge V_-\wedge V_3||}\iota_{V_+\wedge V_-\wedge V_3}\nu~~.
\een

\paragraph{Proposition.} 
We have $\cD|_\cU=(\ker \theta_\perp)\cap \hcD|_\cU$, i.e. the
normalized vector field $\bn^\sharp\in \Gamma(\cU,\hcD)$ is everywhere
orthogonal to $\cD|_\cU$ inside $\hcD|_\cU$, where $~^\sharp$ denotes
the musical isomorphism of $(\hM,\hg)$. Moreover, we have:
\ben
\label{norm_theta0}
||\theta_\perp||=\frac{||V_+\wedge V_-\wedge V_3||}{||\hV_+\wedge \hV_-\wedge \hV_3||}
\een
and:
\ben
\label{nuhnu0}
\nu_\perp=-\iota_\bn\hnu_\perp~~.
\een
Furthermore, we have:
\ben
\label{theta0}
\theta_\parallel=\frac{b_+}{1-\beta^2}\hV_++\frac{b_-}{\beta^2}\hV_-+\frac{b_3}{\beta^2}\hV_3~
~(\mathrm{on}~\cU)
\een
and:
\ben
\label{theta0norm}
||\theta_\perp||^2=1-\frac{b_+^2}{1-\beta^2}-\frac{\rho^2}{\beta^2}~~(\mathrm{on}~\cU)~~,
\een
where $\rho$ was defined in \eqref{rho_def}. 

\

\noindent{\bf Proof.} 
Since $\cD\subset TM$, we have $\cD|_\cU \subset \ker \theta|_\cU$ and
hence $\theta$ vanishes on $\cD|_\cU$. Since $\theta_\parallel$ is a
linear combination of $\hV_1|_\cU,\hV_2|_\cU$ and $\hV_3|_\cU$ and
since $\hcD=\cap_{i=1}^3\ker \hV_i$, we have $\hcD|_\cU \subset
\ker\theta_\parallel$ and hence $\theta_\parallel$ vanishes on
$\hcD|_\cU$ and thus also on $\cD|_\cU\subset \hcD|_\cU$. Using
relation \eqref{theta_dec}, the fact that $\theta|_\cU$ and
$\theta_\parallel$ vanish on $\cD|_\cU$ implies that $\theta_\perp$
vanishes on $\cD|_\cU$ and hence that $\theta_\perp^\sharp$ and
$\bn^\sharp$ are orthogonal to $\cD|_\cU$.  Relations \eqref{nuperp}
and \eqref{hat_nu} give:
\ben
\label{nuperprel}
\nu_\perp=-\frac{1}{||V_+\wedge V_-\wedge V_3||}\iota_{V_+\wedge V_-\wedge V_3\wedge \theta}\hnu|_\cU
=\frac{||\hV_+\wedge \hV_-\wedge \hV_3||}{||V_+\wedge V_-\wedge V_3||}\iota_{\theta_\perp}\hnu_\perp~~,
\een
where in the second equality we used the relation \eqref{hnuperp} and
the equality:
\be
V_+\wedge V_-\wedge V_3\wedge \theta=\hV_+\wedge \hV_-\wedge \hV_3\wedge \theta=-\theta_\perp\wedge \hV_+\wedge \hV_-\wedge \hV_3 ~~,
\ee
which follows from the decompositions \eqref{hVdef} and
\eqref{theta_dec} upon noticing that $\hV_+\wedge \hV_-\wedge
\hV_3\wedge \theta_\parallel=0$. Relations \eqref{norm_theta0} and
\eqref{nuhnu0} now follow from \eqref{nuperprel} upon noticing that
$||\nu_\perp||=||\hnu_\perp||=1$ and
$||\iota_{\theta_\perp}\hnu_\perp||=||\theta_\perp||||\hnu_\perp||=||\theta_\perp||$.
The decomposition \eqref{theta_dec} means that $\theta_\parallel$ is
the projection of $\theta$ onto $\langle
\hV_+|_\cU,\hV_-|_\cU,\hV_3|_\cU\rangle$. Writing
$\theta_\parallel=a_+\hV_+|_\cU+a_-\hV_-|_\cU+a_3\hV_3|_\cU$ with
$a_r\in \cC^\infty(\cU,\R)$, we have $\langle
\theta,\hV_r\rangle|_\cU =\langle \theta_\parallel,\hV_r\rangle|_\cU=
a_r||\hV_r||^2|_\cU$, where we used the fact that $\langle
\hV_r,\hV_s\rangle=||\hV_r|||^2\delta_{rs}$ for all $r,s\in \{+,-,3\}$
(see \eqref{hVsys}). On the other hand, relations \eqref{hVdef} give
$\langle \theta, \hV_r\rangle|_\cU =b_r$. Thus
$a_r=\frac{b_r}{||\hV_r||^2}$ on $\cU$, i.e.:
\be
a_+=\frac{b_+}{||\hV_+||^2}=\frac{b_+}{1-\beta^2}~~,
~~a_-=\frac{b_-}{||\hV_-||^2}=\frac{b_-}{\beta^2}~~,
~~a_3=\frac{b_3}{||\hV_3||^2}=\frac{b_3}{\beta^2}~~(\mathrm{on}~\cU)~~,
\ee
where we used \eqref{betarel}. This immediately gives \eqref{theta0}
and \eqref{theta0norm}. Notice that relation \eqref{theta0norm} can
also be derived from \eqref{norm_theta0} by using the expression for
the Gram determinant $||V_+\wedge V_-\wedge V_3||^2$ given in
\cite[Section 4.2]{msing} and the relation $||\hV_+\wedge \hV_-\wedge
\hV_3||^2=||\hV_+||^2||\hV_-||^2||\hV_3||^2=\beta^4(1-\beta^2)$ (which
follows from \eqref{betarel}). Indeed, we have:
\be
||\theta_\perp||^2= \frac{||V_+\wedge V_-\wedge V_3||^2}{||\hV_+\wedge \hV_-\wedge \hV_3||^2}=-\frac{\beta^2(\beta^4-\beta^2(1-b_+^2+\rho^2)+\rho^2)}{\beta^4(1-\beta^2)}
=\frac{\beta^2(1-\beta^2)-\beta^2b_+^2 -\rho^2(1-\beta^2)}{\beta^2(1-\beta^2)}~~,
\ee 
which recovers \eqref{theta0norm}. $\blacksquare$

\paragraph{Remark.} 
Relations \eqref{theta_dec}, \eqref{n_def}, \eqref{theta0} and
\eqref{theta0norm} give:
\ben
\label{thetaexp}
\theta=_\cU\frac{b_+}{1-\beta^2}\hV_++\frac{b_-}{\beta^2}\hV_-+\frac{b_3}{\beta^2}\hV_3 +\sqrt{1-\frac{b_+^2}{1-\beta^2}-\frac{\rho^2}{\beta^2}}~\bn~~.
\een
Substituting \eqref{hVdef} into this relation allows us to express
$\theta|_\cU$ in terms of $V_+,V_-,V_3$ and $\bn$:
\ben
\label{thetaVn}
\theta=_\cU\frac{\frac{b_+}{1-\beta^2}V_+-\frac{b_-}{\beta^2}V_--\frac{b_3}{\beta^2}V_3}{1-\frac{b_+^2}{1-\beta^2}-\frac{\rho^2}{\beta^2}}+\frac{\bn}{\sqrt{1-\frac{b_+^2}{1-\beta^2}-\frac{\rho^2}{\beta^2}}}~~.
\een
Relation \eqref{thetaexp} should be compared with equation (3.9) of
reference \cite{Palti}, which holds only on the non-transverse locus
$\cN$ (and on its lift to $\hM$). By contrast, equation
\eqref{thetaexp} holds on the generic locus $\cU$, which is contained
in the transverse locus.

\subsection{Relation between $\SU(2)$ and $\SU(3)$ structures}

An $\SU(2)$ structure on the oriented rank five Frobenius distribution
$\cD|_\cU$ which is compatible with the metric $g|_{\cD}$ and with the
orientation of $\cD$ can be described by a normalized one-form
$\balpha\in \Omega^1_{\cU}(\cD)$ and three mutually orthogonal
2-forms $\momega_1,\momega_2,\momega_3\in \Omega^2_{\cU}(\cD)$
satisfying the equations (see \cite{ContiSalamon}):
\ben
\label{su2can}
\begin{split}
& ~\iota_\balpha\momega_k=0\\
&\langle \momega_k,\momega_l\rangle =2\delta_{kl}\\
&~\momega_k\wedge \momega_l=\delta_{kl} \bv~~,
\end{split}~~
\een
where $k,l=1,2, 3$ and $\bv$ is a non-vanishing four-form which
satisfies:
\be
\iota_\balpha\nu_\perp=\frac{1}{2}\bv~~\mathrm{i.e.}~~\balpha\wedge \bv=2\nu_\perp~~.
\ee
Namely, we have $\cD_0|_\cU=\ker \balpha$ and
$(\momega_1,\momega_2,\momega_3)$ is an orthogonal basis of the free
$\cC^\infty(\cU,\R)$-module $\Omega^{2+}_{\cU}(\cD_0)$ of
$\cD_0|_\cU$-longitudinal self-dual 2-forms. As explained in
\cite{ContiSalamon}, this basis can be chosen such that it forms a
positively-oriented frame of the rank three bundle
$\wedge^{2+}\cD_0^\ast$, where the latter is endowed with the
orientation naturally induced from that of $\cD_0$ (which is given by
the volume form $\frac{1}{2}\bv$). 

On the other hand, an $\SU(3)$ structure on the oriented rank six
Frobenius distribution $\hcD|_{\cU}$ which is compatible with the
metric $\hg|_\cD$ and with the orientation of $\hcD$ is determined
\cite{ChiossiSalamon} by an almost complex structure $\bI\in
\Gamma(\cU,\End(\hcD))$ which is compatible with the metric and
orientation of $\hcD$, together with a complex-valued three-form
$\bOmega\in \Omega^2_{\cU}(\hcD)\otimes \C$ which is of unit norm and
has type $(3,0)$ with respect to $\bI$. The almost complex structure
defines a two-form $\bJ\in \Omega^2_{\cU}(\hcD)$ through the relation:
\ben
\label{hIdef}
\bJ(X,Y)\eqdef \hg(X,\bI Y)~~,~~\forall X,Y\in \Gamma(\cU,\hcD)~~
\een
and this form satisfies: 
\ben
\label{hJcube}
\hnu_\perp=_\cU\frac{1}{6}\bJ\wedge\bJ\wedge\bJ~~.
\een
The phase of the normalized (3,0)-form $\bOmega$ is fixed
through the convention:
\ben
\label{hOmegaWedge}
\bOmega\wedge \bar{\bOmega}=_\cU -8 \i \hnu_\perp~~,
\een
Decomposing $\bOmega$ into its real and imaginary parts:
\ben
\label{hOmegaDec}
\bOmega=\bphi+i\brho~~\mathrm{with}~~\bphi,\brho\in \Omega^3_\cU(\hcD)~~,
\een
relation \eqref{hOmegaWedge} amounts to: 
\ben
\label{phirhoWedge}
\bphi\wedge \brho=_\cU 4\hnu_\perp~~.
\een
The following proposition gives the relation between $\SU(3)$
structures on the rank six Frobenius distribution $\hcD|_\cU\eqdef
j^\ast(\hcD)|_\cU$ and $\SU(2)$ structures on its corank one
sub-distribution $\cD|_\cU\subset \hcD|_\cU$.

\paragraph{Proposition.} 
There is a bijective correspondence between $\SU(3)$ structures on
$\hcD|_\cU$ which are compatible with the metric and orientation of
$\hcD|_\cU$ and $\SU(2)$ structures on $\cD|_\cU$ which are compatible with
the metric and orientation of $\cD|_\cU$.  This correspondence is
given as follows, where $\bn$ was defined in \eqref{n_def}:
\begin{enumerate}[(a)]
\itemsep 0.0em
\item Given a metric- and orientation-compatible $\SU(3)$ structure on
  $\hcD|_\cU$ with 2-form $\bJ\in \Omega^2_\cU(\hcD)$ and complex
  3-form $\bOmega\in \Omega^3_\cU(\hcD)\otimes \C$, the following
  formulas give the canonical forms defining the corresponding metric-
  and orientation-compatible $\SU(2)$ structure on $\cD|_\cU$, where
  $\i$ is the imaginary unit:
\beqa
&&\balpha=-\iota_\bn\bJ|_\cD \in \Omega^1_\cU(\cD)~~,~~
\momega_1=\bJ|_\cD\in \Omega^2_\cU(\cD)~~,~~\\
&&\momega_2+\i\momega_3=-\i~ \iota_\bn\bOmega|_\cD \in \Omega^2_\cU(\cD)~~.
\eeqa
\item Given a metric- and orientation-compatible $\SU(2)$ structure on
  $\cD|_\cU$ which is defined by the canonical forms $\balpha\in
  \Omega^1_\cU(\cD)$ and $\momega_k\in \Omega^2_\cU(\cD)$ (where
  $k=1,2,3$), the following forms define the corresponding metric- and
  orientation-compatible $\SU(3)$ structure on $\hcD|_\cU$:
\beqan
\label{momega}
&&\bJ=\momega_1+\balpha\wedge \bn\in \Omega^2_\cU(\hcD)~~,~~\\
\label{Omega}
&&\bOmega=(\momega_2+\i\momega_3)\wedge (\balpha+\i \bn)\in \Omega^3_\cU(\hcD)\otimes \C~~.
\eeqan
\end{enumerate}

\noindent{\bf Proof.} This is an obvious adaptation of
\cite[Proposition 1.4]{ContiSalamon} to the case of Frobenius
distributions. Notice that the signs agree with
our choices of orientation. Indeed, we have:
\be
\balpha\wedge \bv=\balpha\wedge \momega_1\wedge \momega_1=_\cU-(\iota_\bn\bJ) \wedge \bJ\wedge\bJ|_\cD=
-\frac{1}{3}\iota_\bn (\bJ\wedge \bJ\wedge \bJ)|_\cD=_\cU -2 \iota_\bn \hnu_\perp=2\nu_\perp~~.
\ee
$\blacksquare$

\subsection{Recovering the $\SU(2)$ structure on the generic locus of $M$}

Reference \cite{Palti} constructs an $\SU(3)$ structure on the rank
six Frobenius distribution which is obtained by restricting $\hcD$ to
the open subset $\hcU\subset \hM$, a set which (by the results of
Section \ref{sec:t}) contains the $\pi_1^{-1}(\cU)$ of the generic
locus. This $\SU(3)$ structure is described in loc. cit through
certain differential forms denoted there by:
\be
K\in \Omega^2_{\hcU}(\hcD) ~~\mathrm{and}~~\varphi,\rho \in \Omega^3_{\hcU}(\hcD)~~.
\ee
As shown in Appendix \ref{app:P}, the canonically-normalized forms of
that $\SU(3)$ structure are given by:
\beqan
\label{hJhOmegaP}
&&\hJ=_\cU\frac{1}{\sqrt{1-\beta^2}}\left[K-\frac{1}{\beta^2}(\hV_-\wedge\hV_3)\right] \in \Omega^2_{\hcU}(\hcD)\nn\\
&&\hOmega=\hphi+\i \hrho\in \Omega^3_{\hcU}(\hcD)\otimes \C~~,
\eeqan
where: 
\ben
\label{hphirhoP}
\hphi\eqdef \frac{1}{\sqrt{1-\halpha}}\varphi=\frac{1}{\hbeta\sqrt{2}}\varphi~~,~~\hrho\eqdef \frac{1}{\sqrt{1-\halpha}}\sqrt{\frac{2}{1+\halpha}}\rho=\frac{1}{\hbeta\sqrt{2(1-\hbeta^2)}}\rho~~.
\een
Restricting everything to the subset $\cU\equiv j(\cU)\subset
\hcU\subset M$, we obtain an $\SU(3)$ structure on the restricted
Frobenius distribution $\hcD|_\cU$, whose canonically-normalized forms
are given by:
\be
\bJ=\hJ|_\cU~~,~~\bOmega=\hOmega|_\cU~~.
\ee
By definition, the 1-form $\theta_\perp\in \Omega^1(\cU)$ is the
component of $\theta|_\cU$ which is orthogonal to the sub-bundle
$\langle \hV_+|_\cU,\hV_-|_\cU,\hV_3|_\cU \rangle$ of $T^\ast
\hM|_\cU$ generated by the 1-forms $\hV_+|_\cU,\hV_-|_\cU$ and
$\hV_3|_\cU$. Hence the 1-form $\bn$ (which is defined through
\eqref{n_def}) is also orthogonal to this sub-bundle and thus
$\iota_\bn \hV_k=0$ for all $k$. On the other hand, we have
$\cD|_\cU=\ker \theta\cap \hcD|_\cU=\ker \bn \cap \hcD|_\cU\subset
\ker \bn$ and hence $\momega_k$ and $\balpha$ (which are longitudinal
to the Frobenius distribution $\cD|_\cU$) are also orthogonal to
$\bn$. These observations show that we have the relations:
\ben
\label{nperp}
\iota_\bn \hV_k=\iota_\bn\momega_k=\iota_\bn\balpha=0~~,~~\forall k=1,2,3~~.
\een
Using \eqref{nperp}, relation \eqref{momega} implies:
\be
\balpha=-\iota_\bn \bJ=-\iota_\bn \hJ|_\cU~~.
\ee
Substituting the first relation of \eqref{hJhOmegaP}, this gives:
\ben
\label{rel1}
\balpha=-\frac{1}{\sqrt{1-\beta^2}}\iota_\bn K|_\cU~~.
\een
Now \eqref{momega} and \eqref{hJhOmegaP} give: 
\ben
\label{rel2}
\momega_1=\bJ+\bn\wedge\balpha=\frac{1}{\sqrt{1-\beta^2}}\left[K-\bn\wedge (\iota_\bn K)
  -\frac{1}{\beta^2}\hV_-\wedge\hV_3 \right]|_\cU~~.
\een
Relation \eqref{Omega} expands as:
\be
\bOmega=(\momega_2+\i\momega_3)\wedge\balpha+(\i\momega_2-\momega_3)\wedge \bn=
(\momega_2\wedge\balpha-\momega_3\wedge\bn)+\i(\momega_3\wedge\balpha+\momega_2\wedge\bn)~~.
\ee
Comparing this with the second relation in \eqref{hJhOmegaP} gives:
\be
\hvarphi|_\cU\eqdef\bphi=\balpha\wedge \momega_2-\bn\wedge\momega_3~~,~~
\hrho|_\cU\eqdef\brho=\balpha\wedge \momega_3+\bn\wedge\momega_2~.
\ee
Using \eqref{hphirhoP} and \eqref{nperp}, these equations imply: 
\ben
\label{rel3}
\momega_2=\iota_\bn\hrho|_\cU=\frac{1}{\beta\sqrt{2(1-\beta^2)}}\iota_\bn \rho|_\cU~~,~~
\momega_3=-\iota_\bn\hphi|_\cU=-\frac{1}{\beta\sqrt{2}}\iota_\bn \varphi|_\cU~~.
\een
Relations \eqref{rel1}, \eqref{rel2} and \eqref{rel3} express the
defining forms of the $\SU(2)$ structure on the generic locus
$\cU\subset M$ in terms of the defining forms of the $\SU(3)$
structure which exists on the locus $\hcU\subset \hM$.

\section{Conclusions} 

We analyzed the stabilizer stratifications of internal eight-manifolds
$M$ which can arise in $\cN=2$ M-theory flux compactifications down to
three dimensions using the formalism based on the auxiliary
nine-manifold $\hM\eqdef M\times S^1$, which can be viewed as a
trivial circle bundle over $M$ with projection $\pi_1$. We showed how
the complicated stratified G-structure of $M$ which was uncovered in
\cite{msing} relates to the much simpler stratified G-structure of
$\hM$. The increased complexity of the former arises from the fact
that the cosmooth generalized distribution $\hcD$ whose rank
determines the stabilizer stratification of $\hM$ may have pointwise
transverse or non-transverse intersection with the $\pi_1$-pull-back
of the tangent bundle of $M$. We also gave an explicit construction of
the defining forms of the $\SU(2)$ structure which exists on the
generic locus $\cU\subset M$ in terms of the defining forms of the
$\SU(3)$ structure which exists on the locus $\hcU\subset \hM$.

\acknowledgments{The work of E.M.B. was partly supported by the
  strategic grant POSDRU/159/1.5/S/133255, Project ID 133255 (2014),
  co-financed by the European Social Fund within the Sectorial
  Operational Program Human Resources Development 2007 -- 2013 and
  partly by the CNCS-UEFISCDI project PN-II-ID-PCE 121/2011.  The work
  of C.I.L was supported by the research grants IBS-R003-G1 and
  IBS-R003-S1.}

\appendix

\section{Canonically-normalized forms of the $\SU(3)$ structure on $\hM$}
\label{app:P}

Reference \cite{Palti} constructs a two-form $J\in
\Omega^2_{\hcU}(\hcD)$ given by equation (2.29) of loc. cit. When
translated into our notations, that equation amounts to:
\ben
\label{hJP}
J=_{\hcU} K-\frac{2}{1-\halpha}\hV_-\wedge\hV_3=K-\frac{1}{\hbeta^2}\hV_-\wedge\hV_3~~,
\een
where (as in \cite{Palti}):
\be
K\eqdef \frac{1}{2}\hcB(\hxi_1,
\hgamma_{m_1m_2}\hxi_2)\he^{m_1}\wedge \he^{m_2}\in \Omega^2(\hM)~~.
\ee
To arrive at \eqref{hJP}, we used relation \eqref{hatalpha} and the
fact that $\hV_\pm^\mathrm{here}=\frac{1}{2}V_\pm^\mathrm{there}$. By
the construction given in \cite{Palti} (see the derivation of
eq. (2.29) of loc. cit. and the discussion preceding it), the 2-form
$J$ coincides with the orthogonal projection of $K$ onto
$\Omega^2(\hcD)\subset \Omega^2(\hM)$. We thus have
$\iota_{\hV_k}J=0$ for all $k=1,2,3$ and hence $J$ is a two-form
defined on $\hM$ which is longitudinal to the distribution
$\hcD$. Define $I\in \Gamma(\hcU,\End(T\hM))$ through:
\ben
\label{Idef}
J(X,Y)\eqdef \hg(X, I Y)~~,~~\forall X,Y\in \Gamma(\hcU,T\hM)~~.
\een
In a local frame $\he_m$ of $\hM$ defined over an open subset $U\subset
\hM$, we have $I\he_n=I_n^{~~p}\he_p$ and $J(\he_m,\he_n)=J_{mn}=-J_{nm}$, hence
\eqref{Idef} becomes:
\be
J_{mn}=-J_{nm}=_UI_n^{~~p}\hg_{pm}~~,
\ee
where $\hg_{mp}=\hg_{pm}=\hg(\he_m,\he_p)$. Thus $I_m^{~~n}=-J_{mp}\hg^{pn}=-J_m^{~~n}$ and
$I^2(\he_m)=I_m^{~~n}I(\he_n)=I_m^{~~n}I_n^{~~p} \he_p=J_{mn}J^{np}\he_p$.
Hence equation (2.36) of \cite{Palti} implies: 
\ben
I^2|_{\hcD}=_{\hcU}-\frac{1+\halpha}{2}\id_{\hcD} ~~.
\een
Therefore, the quantity:
\ben
\label{hIP}
\hI\eqdef \sqrt{\frac{2}{1+\halpha}} I=\frac{1}{\sqrt{1-\hbeta^2}}I\in \Gamma(\hcU,T\hM)
\een
satisfies $\hI^2|_{\hcD}=_{\hcU}-\id_{\hcD}$ and hence it gives an
almost complex structure on the rank six Frobenius distribution which
is obtained by restricting $\hcD$ to $\hcU$. The two-form associated
to this almost complex structure is given by:
\ben
\label{hJ}
\hJ=\sqrt{\frac{2}{1+\halpha}}J=\frac{1}{\sqrt{1-\hbeta^2}}J=
\frac{1}{\sqrt{1-\hbeta^2}}\left[K-\frac{1}{\hbeta^2}\hV_-\wedge\hV_3\right]\in \Omega^2_{\hcU}(\hcD)
\een
and satisfies the analogue of the normalization condition \eqref{hJcube} (on $\hcU$)
by virtue of equation\footnote{Notice that there is a typo in
  \cite[eq. (2.40)]{Palti} in that the right hand side of that
  equation should equal $\frac{3(1+\alpha)}{4(1-\alpha)}\ast
  (V_+\wedge V_-\wedge V_3)$ (in the notations of loc. cit.). With
  this correction, that equation is equivalent in our notations with
  $J\wedge J\wedge J=\frac{3(1+\alpha)}{(1-\halpha)} {\hat \ast}
  (\hV_+\wedge \hV_-\wedge \hV_3)$, where we used the fact that
  $\hV_\pm^\mathrm{here}=\frac{1}{2}V_\pm^\mathrm{there}$. Relation
  \eqref{GramhV} implies ${\hat \ast} (\hV_+\wedge \hV_-\wedge
  \hV_3)=(1-\halpha)\sqrt{\frac{1+\halpha}{2}}\hnu_\perp$ and hence
  $J\wedge J\wedge J=6\left(\frac{1+\halpha}{2}\right)^{3/2}
  \hnu_\perp$ i.e. $\hJ\wedge \hJ\wedge \hJ=6\hnu_\perp$. Here ${\hat
    \ast}$ denotes the Hodge operator of $(\hM,\hg)$.} (2.40) of
\cite{Palti}. Loc. cit. also constructs two real 3-forms $\varphi$ and
$\rho$ on $\hcU$ which are orthogonal to $\hV_+,\hV_-$ and $\hV_3$ on
$\hcU$ and hence belong to the space $\Omega^3_{\hcU}(\hcD)$ (see page
10 of \cite{Palti}). These forms satisfy relation (2.39) of
\cite{Palti}, which in our notations reads:
\ben
\label{phirjoWP}
\varphi\wedge \rho=_{\hcU} 4 {\hat \ast} (\hV_+\wedge \hV_-\wedge \hV_3)=4(1-\halpha)\sqrt{\frac{1+\halpha}{2}}\hnu_\perp~~,
\een
where we used \eqref{GramhV} and the fact that
$\hV_\pm^\mathrm{here}=\frac{1}{2}V_\pm^\mathrm{there}$. Defining:
\beqan
\label{hphihrho}
&& \hphi\eqdef \frac{1}{\sqrt{1-\halpha}}\varphi=\frac{1}{\hbeta\sqrt{2}}\varphi\nn\\
&&\hrho\eqdef \frac{1}{\sqrt{1-\halpha}}\sqrt{\frac{2}{1+\halpha}}\rho=\frac{1}{\hbeta\sqrt{2(1-\hbeta^2)}}\rho~~,
\eeqan
relation \eqref{phirjoWP} reduces to the analogue of
\eqref{phirhoWedge}, which holds on $\hcU$. Loc. cit. also defines a
complex-valued 3-form $\Omega\in \Omega^3_{\hcU}(\hcD)\otimes \C$ 
through \cite[eq. (2.41)]{Palti}, which reads:
\be
\Omega\eqdef \varphi+\i\sqrt{\frac{2}{1+\halpha}}\rho~~.
\ee
Defining: 
\ben
\label{hOmega}
\hOmega\eqdef \frac{1}{\sqrt{1-\halpha}}\Omega=\frac{1}{\hbeta\sqrt{2}}\Omega=\hphi+\i \hrho~~,
\een
relation (2.42) of \cite{Palti} becomes the condition that $\hOmega$
is $\hI$-pseudoholomorphic: 
\be 
\hI_{(1)}\hOmega=-\i~\hOmega~~(\mathrm{i.e.}~\hI_{(1)}\hphi=\hrho)~~, 
\ee
where $\hI_{(1)}$ denotes the action of $\hI$ on the first ``slot'' of
$\hOmega$. On the other hand, the analogue of relation
\eqref{phirhoWedge} shows that $\hOmega$ satisfies the analogue of
\eqref{hOmegaWedge} on $\hcU$. Combining everything, we conclude that
$\hJ$ and $\hOmega$ are the canonically-normalized forms of the
$\SU(3)$ structure which was constructed in \cite{Palti} on the rank
six Frobenius distribution obtained by restricting $\hcD$ to the locus
$\hcU\subset \hM$.



\begin{thebibliography}{100}
\bibitem{sugra11}{E.~Cremmer, B.~Julia, J.~Scherk, {\em
    Supergravity Theory in Eleven-Dimensions}, Phys. Lett. {\bf B 76}
  (1978) 409.}
\bibitem{BeckerCY}{K.~Becker, M.~Becker, ``M theory on eight
  manifolds,'' Nucl. Phys. {\bf B 477} (1996) 155.}
\bibitem{Palti}{C.~Condeescu, A.~Micu, E.~Palti, {\em M-theory
    Compactifications to Three Dimensions with M2-brane Potentials},
  JHEP {\bf 04} (2014) 026.}
\bibitem{ga2}{C.~I.~Lazaroiu, E.~M.~Babalic, {\em Geometric algebra
    techniques in flux compactifications (II)}, JHEP {\bf 06} (2013)
  054.}
\bibitem{msing}{E.~M.~Babalic, C.~I.~Lazaroiu, {\em The landscape of
    G-structures in eight-manifold compactifications of M-theory}, arXiv:1505.02270 [hep-th].}
\bibitem{Whitney}{H. Whitney, {\em Elementary Structure of real
    algebraic varieties}, Ann. Math. {\bf 66} (1957), 545--556.}
\bibitem{Gibson}{C.~G.~Gibson, K.~Wirthmuller, A.~A.~Du Plessis, E.~J.~N.~Looijenga, 
{\em Topological Stability of Smooth Mappings}, L. N. M {\bf 552}, Springer-Verlag, New York, 1976.}
\bibitem{BCR}{J.~Bochnak, M.~Coste, M.~F.~Roy, {\em Real algebraic
    geometry}, Ergebnisse der Mathematik und ihrer
  Grenzgebiete. 3. Folge, vol. {\bf 36}, Spinger, 1998.}
\bibitem{BPR}{S.~Basu, R.~Pollack, M.~F.~Roy, {\em Algorithms in Real
    Algebraic Geometry}, Algorithms and Computation in Mathematics
  Volume {\bf 10}, Springer, 2006.}  
\bibitem{MartelliSparks}{D.~Martelli and J.~Sparks, {\em G-structures,
    fluxes and calibrations in M-theory}, Phys.  Rev.  {\bf D 68}
  (2003) 085--014.}
\bibitem{Tsimpis}{D.~Tsimpis, {\em M-theory on eight-manifolds
    revisited: N = 1 supersymmetry and generalized $\Spin(7)$
    structures}, JHEP {\bf 04} (2006) 027.}
\bibitem{g2}{Calin Lazaroiu, Mirela Babalic, {\em Foliated
    eight-manifolds for M-theory compactification}, JHEP {\bf 01} (2015) 140.}
\bibitem{g2s}{Calin Lazaroiu, Mirela Babalic, {\em Singular foliations 
    for M-theory compactification}, JHEP {\bf 03} (2015) 116.}
\bibitem{Becker1}{K.~Becker, {\em A Note on compactifications on
    $\Spin(7)$ -- holonomy manifolds}, JHEP {\bf 05} (2001) 003.}
\bibitem{Becker2}{M.~Becker, D.~Constantin, S.~J.~Gates, Jr.,
  W.~D.~Linch, III, W.~Merrell, J.~Phillips, {\em M theory on
    $\Spin(7)$ manifolds, fluxes and 3-D, N=1 supergravity},
  Nucl. Phys. {\bf B 683} (2004) 67.}
\bibitem{Constantin}{D.~Constantin, {\em M-theory vacua from Warped
    compactifications on $\Spin(7)$ holonomy manifolds},
  Fortsch. Phys. {\bf 53} (2005), 1272--1329.}
\bibitem{Drager}{L.~D.~Drager, J.~M.~Lee, E.~Park, K.~Richardson, {\em
    Smooth distributions are finitely generated}, Ann. Global
  Anal. Geom. {\bf 41} (2012) 3, 357--369.}
\bibitem{Freeman}{M.~Freeman, {\em Fully integrable Pfaffian systems},
  Ann. Math. {\bf 119} (1984), 465--510.}
\bibitem{BulloLewis}{F.~Bullo, A.~Lewis, {\em Geometric Control of
    Mechanical Systems}, Texts in Applied Mathematics {\bf 49},
  Springer, 2004.}
\bibitem{Michor}{P.~W.~Michor, {\em Topics in Differential Geometry},
  Graduate Studies in Mathematics {\bf 93}, 2008.}
\bibitem{Ratiu}{J.~E.~Marsden, T. S. Ratiu, internet supplement for
  {\em Introduction to Mechanics and Symmetry}.}
\bibitem{FinoTomassini}{A.~Fino, A. ~Tomassini, {\em Generalized
    $\G_2$-manifolds and $\SU(3)$-structures}, Internat. J. Math. {\bf
    19} (2008) 10, 1147--1165.}
\bibitem{ChiossiSalamon}{S.~Chiossi, S.~Salamon, {\em The intrinsic
    torsion of $\SU(3)$ and $\G_2$ structures}, Differential Geometry,
  Valencia 2001, World Sci. Publishing, 2002, pp. 115--133.}
\bibitem{CabreraSU3}{F.~M.~Cabrera, {\em $\SU(3)$-Structures on
    Hypersurfaces of Manifolds With $\G_2$-Structure}, Monatshefte fur
  Mathematik {\bf 148} (2006) 1, 29--50.}
\bibitem{Knotes}{S.~Karigiannis, {\em Some Notes on $\G_2$ and
    $\Spin(7)$ Geometry}, Recent Advances in Geometric Analysis,
  Advanced Lectures in Mathematics {\bf 11} (2010) 129--146.}
\bibitem{ContiSalamon}{D.~Conti and S. Salamon, {\em Generalized
    Killing spinors in dimension 5}, Trans. AMS. {\bf 359} (2007)
  5319--5343.}
\bibitem{Bedulli}{L.~Bedulli, L.~Vezzoni, {\em Torsion of
    $\SU(2)$-structures and Ricci curvature in dimension 5},
  Diff. Geom. Appl. {\bf 27} (2009) 1, 85--99.}
\bibitem{AC1}{D.~V.~Alekseevsky, V.~Cortes, {\em Classification of
    $N$-(super)-extended Poincare algebras and bilinear invariants of
    the spinor representation of $\Spin(p,q)$}, Commun. Math. Phys. {\bf 183} (1997) 3, 477--510.}
\bibitem{AC2}{D.~V.~Alekseevsky, V.~Cortes, C.~Devchand, A.~V.~Proyen,
  {\em Polyvector Super-Poincare Algebras}, Commun. Math. Phys. {\bf
    253} (2005) 2, 385--422.}
\bibitem{Friedrich}{T.~Friedrich, {\em Weak $\Spin(9)$-Structures on
    16-dimensional Riemannian Manifolds}, Asian Journal of Mathematics
  {\bf 5} (2001), 129--160.}
\end{thebibliography}
\end{document}